\documentclass[journal,twocolumn]{IEEEtran}
\usepackage[T1]{fontenc}
\usepackage{amsmath,amsfonts}
\usepackage{algorithmic}
\usepackage[linesnumbered,ruled,vlined]{algorithm2e}
\usepackage{array}
\allowdisplaybreaks

\usepackage{caption}
\usepackage{xcolor}
\usepackage[justification=centering]{caption}
\usepackage{epsfig,amssymb,subfigure,bm,dsfont}
\usepackage{textcomp}
\usepackage{stfloats}
\usepackage{url}
\usepackage{verbatim}
\usepackage{graphicx}
\usepackage{cite}
\usepackage{multirow}
\usepackage{bbm}
\usepackage{amssymb}
\newtheorem{assumption}{\textbf{Assumption}}
\newtheorem{theorem}{\textbf{Theorem}}

\newtheorem{definition}{\textbf{Definition}}
\newtheorem{lemma}{\textbf{Lemma}}

\newcommand{\tabincell}[2]{\begin{tabular}{@{}#1@{}}#2\end{tabular}}

\renewcommand{\algorithmicrequire}
{\textbf{Require:}}
\renewcommand{\algorithmicensure}
{\textbf{Ensure:}}
\renewcommand{\algorithmicreturn}
{\textbf{Return}}

\begin{document}

\title{Low-latency Federated Learning with DNN Partition in Distributed Industrial IoT Networks}

\author{Xiumei~Deng,~\IEEEmembership{Student Member, IEEE},~Jun~Li,~\IEEEmembership{Senior Member, IEEE},~Chuan~Ma,~\IEEEmembership{Member, IEEE},\\~Kang~Wei,~\IEEEmembership{Graduate Student Member, IEEE},~Long~Shi,~\IEEEmembership{Senior Member, IEEE},\\~Ming~Ding,~\IEEEmembership{Senior Member, IEEE},~and~Wen~Chen,~\IEEEmembership{Senior Member, IEEE}
\thanks{X. Deng, J. Li, K. Wei and L. Shi are with School of Electronic and Optical Engineering, Nanjing University of Science and Technology, Nanjing, 210094, China. E-mail: \{xiumeideng, jun.li, kang.wei\}@njust.edu.cn, slong1007@gmail.com.}
\thanks{C. Ma is with Zhejiang Lab, Hangzhou, China. He is also with Nanjing University of Science and Technology, and Key Laboratory of Computer Network and Information Integration (Southeast University), Ministry of Education. E-mail: chuan.ma@zhejianglab.edu.cn.}
\thanks{M. Ding is with Data61, CSIRO, Sydney, NSW 2015, Australia. E-mail: ming.ding@data61.csiro.au.}
\thanks{W. Chen is with Department of Electronics Engineering, Shanghai Jiao Tong University, Shanghai 200240, China. E-mail: wenchen@sjtu.edu.cn.}
\thanks{Manuscript received May 16, 2022; revised September 06, 2022.}}

\markboth{Accepted by IEEE Journal on Selected Areas in Communications}%
{Shell \MakeLowercase{\textit{et al.}}: A Sample Article Using IEEEtran.cls for IEEE Journals}


\maketitle
\begin{abstract}
Federated Learning (FL) empowers Industrial Internet of Things (IIoT) with distributed intelligence of industrial automation thanks to its capability of distributed machine learning without any raw data exchange. However, it is rather challenging for lightweight IIoT devices to perform computation-intensive local model training over large-scale deep neural networks (DNNs). Driven by this issue, we develop a communication-computation efficient FL framework for resource-limited IIoT networks that integrates DNN partition technique into the standard FL mechanism, wherein IIoT devices perform local model training over the bottom layers of the objective DNN, and offload the top layers to the edge gateway side. Considering imbalanced data distribution, we derive the device-specific participation rate to involve the devices with better data distribution in more communication rounds. Upon deriving the device-specific participation rate, we propose to minimize the training delay under the constraints of device-specific participation rate, energy consumption and memory usage. To this end, we formulate a joint optimization problem of device scheduling and resource allocation (i.e. DNN partition point, channel assignment, transmit power, and computation frequency), and solve the long-term min-max mixed integer non-linear programming based on the Lyapunov technique. In particular, the proposed dynamic device scheduling and resource allocation (DDSRA) algorithm can achieve a trade-off to balance the training delay minimization and FL performance. We also provide the FL convergence bound for the DDSRA algorithm with both convex and non-convex settings. Experimental results demonstrate the derived device-specific participation rate in terms of feasibility, and show that the DDSRA algorithm outperforms baselines in terms of test accuracy and convergence time.
\end{abstract}


\section{Introduction}
\IEEEPARstart{R}{ecent} advances in artificial intelligence (AI) and communication technologies along with wide deployment of the Industrial Internet of Things (IIoT) are leading us to Industry 4.0\cite{DBLP:journals/tii/Zhou0DH21,yyy}. With the proliferation of modern sensors and controllers, massive data has been collected and analyzed to derive intelligence, which transforms traditional industrial manufacturing to modernization and intelligence. Since raw industrial data often involves sensitive information, privacy is of the essence in industrial big data analysis. In traditional machine learning (ML), a large amount of raw data is collected from IIoT devices and centralized in a third party, which can lead to privacy leakage. Therefore, it is of crucial significance to process industrial raw data locally for the sake of privacy protection. As an emerging AI technology, federated learning (FL) has been regarded as a promising solution to privacy-preserving intelligent IIoT applications\cite{DBLP:journals/wc/NguyenDPSLNP21}. In a FL network, the centralized server aggregates the trained local ML models transmitted from the distributed devices. Compared with traditional ML, FL achieves a global ML model without any raw data exchange, thereby significantly reducing the communication overhead and promoting the privacy of each device. In this context, FL is widely applicable to a variety of IIoT scenarios wherein the local data samples possessed by every single device are insufficient to train an efficient ML model.

Wireless communication is of the essence in IIoT scenarios because it enables seamless, pervasive, and scalable connectivity among distributed devices without any cabled connections\cite{DBLP:journals/wc/ZhouSDDN21}. The total cost of deploying cabled communication can be relatively high in an industrial environment, especially when it requires equipment shutdown and pauses manufacturing lines. Therefore, many factories upgrade the existing equipment with industrial wireless components for additional intelligent applications (e.g., fault diagnosis and safety early warning), instead of deploying cables\cite{DBLP:journals/jsac/ZhangYWPZZS21}. However, due to the limited communication resources (e.g., bandwidth), FL in IIoT systems suffers from inter-channel interference and prolonged transmission latency in the training process. Since IIoT systems demand the timely completion of each processing step during the manufacturing process, the communication overhead can be a bottleneck in FL-enabled IIoT systems. In addition, since IIoT devices are typically battery-operated, low-power consumption is vital to preserve battery life\cite{sisinni2021wireless}. To this end, joint communication and energy resource allocation are of considerable significance to improve FL efficiency in wireless IIoT networks.

From the perspective of resource allocation, many recent studies have focused on how to achieve energy-efficient or communication-efficient FL. Authors in \cite{DBLP:journals/tcom/WaduSB21} derive the training loss gap between FL and centralized ML for a given duration of communication rounds, and propose a device scheduling and resource allocation policy to enhance the FL performance by predicting channel state information. Based on deep multi-agent reinforcement learning, the works in \cite{DBLP:journals/jsac/ZhangYWPZZS21} and \cite{DBLP:conf/icc/ZhangYWPZS21} optimize the device selection as well as communication and computation resource allocation in an online manner to minimize FL loss function under delay and energy consumption constraints. To achieve a communication-efficient FL, \cite{DBLP:journals/tii/GaoZMNJ21,DBLP:journals/twc/XuWC22,9748976,DBLP:journals/iotj/ChenHWZZLH21} minimize the training latency by jointly optimizing communication and computation resource allocation as well as device selection. To evaluate the learning performance of the proposed low-latency FL framework, both FL training loss and learning delay are investigated to characterize the impact of device selection and resource allocation on the FL performance. In addition, to overcome the battery challenge of IIoT devices, the works in \cite{DBLP:journals/tvt/PhamZRHH21,9709639,DBLP:journals/iotj/YuASAD22,9507054,DBLP:conf/icc/RenS0NNW21} propose different algorithms of joint communication and energy resource management, aiming at minimizing the total energy consumption of FL training process or weighted sum of energy consumption, latency and FL training loss.

Although emerging AI endows IIoT the capability to mine efficient knowledge from big data, the state-of-the-art deep neural network (DNN) architectures (e.g., GPT3) demand significant memory and computational resources\cite{9364277}. Considering the huge computational cost of large-scale DNN training, the aforementioned works on communication and computation resource allocation are not adequate to reduce the computational burden on lightweight IIoT devices during the FL training process. To further reduce the computational cost of FL training for IIoT devices, recent works \cite{DBLP:conf/mlicom/ZhangWZ019,DBLP:conf/ics/DemirciF21,DBLP:journals/monet/ZhangZWC20} on DNN partition assisted FL propose to divide the DNN model into two continuous portions, and separately train bottom and top layers of the DNN model at the device and edge server sides. However, these works focus on differentially private data perturbation mechanism designs to preserve the privacy of training data, and adopt predefined DNN partition strategies for all devices regardless of limited and heterogeneous computational resources. Later on, \cite{DBLP:conf/infocom/MohammedJBF20} and \cite{9542866} propose to jointly optimize the partitioning and offloading of DNN inference tasks to reduce the total execution time. However, these works focus on DNN inference instead of DNN training. In fact, it is more challenging to optimize the DNN partition point in FL training process. This is due to the fact that FL demands the devices to synchronously perform the local model training in each communication round, while the DNN inference tasks can be independently executed by the devices. To our best knowledge, our paper is the first attempt to investigate the dynamic DNN partition in FL training process.

In addition, due to diverse computational capacity and memory resource among different IIoT devices, device heterogeneity introduces high training latency or even training failures, resulting in an unsatisfactory quality of experience (QoE) for real-time delay-sensitive applications. Furthermore, data heterogeneity can significantly degrade FL performance in the presence of non-independent and identically distributed (non-IID) data distribution\cite{9347706}. To this end, proper participant device selection plays a crucial role in improving FL performance, especially when there exits a limit on the number of participant devices due to the limited communication resources.

To address the aforementioned issues, we propose a communication-computation efficient FL framework for resource-limited IIoT networks that integrates the dynamic DNN partition technique into the standard FL mechanism. Considering limited and heterogeneous communication, energy, and memory resources as well as imbalanced data distribution, we propose a dynamic device scheduling and resource allocation policy to minimize the training latency while guaranteeing FL performance. Our contributions are summarized as follows:
\begin{enumerate}
	\item By integrating DNN partition with FL, we propose a two-tier FL framework for IIoT networks wherein the devices hold the private datasets, perform the local model training over the bottom layers of the objective DNN, and offload the top layers to the edge gateway side at the middle tier. The roles of edge gateways include local model training of top DNN portion, device-level model combination, shop-floor-scale model aggregation, and model transmission. The base station (BS) at the top tier performs global model aggregation, and transfers scheduling policy information.		
	\item According to the forward and backward propagation, we derive the universal formulas for evaluating the layer-level memory usage and floating-point operation counts (FLOPs) based on the hyper-parameters of the DNN structure (e.g., filter size in convolution layer). As such, we propose a layer-level calculation model for training delay, memory usage and energy consumption in our two-tier FL framework.
	\item We derive a divergence bound to analyze the impact of local dataset size and data distribution on FL training performance, and then develop the device-specific participation rate linked to the model performance. To achieve a low-latency FL, we formulate a joint dynamic optimization problem of the device scheduling and resource allocation (i.e., channel assignment, DNN partition point, transmit power and computation frequency) under the constraints of device-specific participation rate, energy consumption and memory usage. The objective of this optimization problem is to minimize the training latency while guaranteeing the learning performance of FL. To solve this long-term min-max mixed integer non-linear programming (MINLP) problem, we propose a dynamic device scheduling and resource allocation (DDSRA) algorithm to transform the stochastic optimization problem with a time-average device-specific participation rate constraint into a deterministic min-max MINLP problem based on the Lyapunov technique. Then, we solve the deterministic problem by the block coordinate descent method and bisection method in each communication round.
	\item We conduct a performance analysis of the proposed DDSRA algorithm to verify its asymptotic optimality. A trade-off of [$\mathcal{O}(1/V)$, $\mathcal{O}(\sqrt{V})$] is characterized between the FL training latency minimization and the degree of which the participation rate constraint is satisfied with a control parameter $V$. This trade-off indicates that the minimization of the training latency and FL performance can be balanced by adjusting $V$. Furthermore, we provide the FL convergence bound for the DDSRA algorithm with both convex and non-convex settings. Our developed bound reveals that the FL convergence rate can be improved by increasing the training data size and setting a higher participation rate for the important devices with better data distribution.
	\item Experimental results are provided to demonstrate the derived device-specific participation rate in terms of feasibility. Moreover, we analyze the participation rate of each device under the proposed DDSRA algorithm, and the experimental results show that the DDSRA algorithm outperforms the baselines in terms of learning accuracy and convergence time.
\end{enumerate}

The remainder of this paper is organized as follows. In Section \ref{sec:Preliminaries}, we briefly introduce the basic knowledge of FL and DNN partition technique. In Section \ref{sec:System Model}, we give the communication-computation efficient FL-enabled IIoT framework and then formulate the stochastic optimization problem. Section \ref{sec:Device-specific measurement of the participation rate} derives the device-specific participation rate, and Section \ref{sec:Dynamic channel assignment and resource allocation algorithm} proposes the DDSRA algorithm. Section \ref{sec:Performance Analysis} produces the performance analysis of the proposed algorithm to verify asymptotic optimality, and studies the FL convergence rate. Then, the experimental results are presented in Section \ref{sec:Experiential Results}. Section \ref{sec:Conclusion} concludes this paper. For ease of reference, Tables I lists the main notations used in this paper.
\begin{table}[!t]
	\captionsetup{labelfont={scriptsize},font={scriptsize}}
	\caption{List of main notations.}
	\label{tab:SummaryofMainNotations}
	\renewcommand{\arraystretch}{1.2}
	\centering
	\scalebox{0.9}{
		{\begin{tabular}{c||l}
				\hline
				\bfseries Notations & \bfseries Descriptions\\
				\hline\hline
				$\mathcal{N}$& Index set of the end devices\\
				\hline
				$\mathcal{M}$& Index set of the edge gateways\\
				\hline
				$\boldsymbol{a}$& Deployment matrix\\
				\hline
				$\mathcal{D}_n$& Local dataset\\
				\hline
				$\mathcal{L}$& Index set of the DNN layers\\
				\hline
				$\mathcal{J}$&Index set of the available channels\\
				\hline
				$l_n(t)$&DNN partition point\\
				\hline
				$\boldsymbol{I}(t)$&Channel assignment matrix\\
				\hline
				$K$&Local iterations\\
				\hline
				$\beta$&Step size\\
				\hline
				$\tilde{D}_n$&Number of sample points\\
				\hline
				$o_l$&\tabincell{l}{FLOPs of the forward propagation \\for each sample point in the $l$-th layer}\\
				\hline
				$o'_l$&\tabincell{l}{FLOPs of the backward propagation \\for each sample point in the $l$-th layer}\\
				\hline
				$\phi_n^\text{D}$&FLOPs per clock cycle of the $n$-th device\\
				\hline
				$\phi_m^\text{G}$&FLOPs per clock cycle of the $m$-th gateway\\
				\hline
				$f_n^\text{D}$&Computation frequency of the $n$-th device\\
				\hline
				$f_{m,n}^\text{G}(t)$&\tabincell{l}{Computation frequency of the $m$-th gateway assigned to the\\ local model training offloaded from the $n$-th device in the $t$-th\\ communication round}\\
				\hline
				$e_n^\text{tra,D}(t)$&\tabincell{l}{Energy consumption of the $n$-th device for local\\ model training in the $t$-th communication round}\\
				\hline
				$v_n^\text{D}$&Effective switched capacitance of the $n$-th device\\
				\hline
				$e_m^\text{tra,G}(t)$&\tabincell{l}{Energy consumption of the $m$-th gateway for \\local model training in the $t$-th communication round}\\
				\hline
				$g_{n,l}$&\tabincell{l}{Memory usage of the $l$-th layer for storing the model parameters \\and intermediate data in the forward and backward propagation}\\
				\hline
				$G_n^\text{D}(t)$&\tabincell{l}{Memory usage for the bottom DNN layers trained\\ at the $n$-th device in the $t$-th communication round}\\
				\hline
				$G_m^\text{G}(t)$&\tabincell{l}{Memory usage for the top DNN layers trained\\ at the $m$-th gateway in the $t$-th communication round}\\
				\hline
				$G_n^\text{D,max}$&Memory size of the $n$-th device\\
				\hline
				$G_m^\text{G,max}$&Memory size of the $m$-th gateway\\
				\hline
				$h_{m,j}^\text{d}(t)$&\tabincell{l}{Downlink channel power gain from the BS to the $m$-th \\gateway via the $j$-th channel in the $t$-th communication round}\\
				\hline
				$B^\text{d}$&Bandwidth of the downlink channel\\
				\hline
				$P^B$&Transmit power of the BS\\
				\hline
				$N_0$&Noise power spectral density\\
				\hline
				$\gamma$&DNN model size\\
				\hline
				$i^\text{d}_{m,j}(t)$&Co-channel interference of the downlink channel\\
				\hline
				$B^\text{u}$&Bandwidth of the uplink channel\\
				\hline
				$P_m(t)$&\tabincell{l}{Transmit power of the $m$-th gateway\\ in the $t$-th communication round}\\
				\hline
				$h^u_{m,j}(t)$&\tabincell{l}{Uplink channel power gain from the $m$-th \\gateway to the BS in the $t$-th communication round}\\
				\hline
				$i^u_{m,j}(t)$&Co-channel interference of the uplink channel\\
				\hline
				$e_m^\text{up}(t)$&\tabincell{l}{Energy consumption of the $m$-th gateway for \\model transmitting in the $t$-th communication round}\\
				\hline
				$E_n^\text{D}(t)$&\tabincell{l}{Energy arrival at the $n$-th device in the $t$-th communication round}\\
				\hline
				$E_m^\text{G}(t)$&\tabincell{l}{Energy arrival at the $m$-th gateway \\in the $t$-th communication round}\\
				\hline
				$e_m^\text{G}(t)$&\tabincell{l}{Total energy consumption of the $m$-th \\gateway in the $t$-th communication round}\\
				\hline
				$\tau(t)$&Total latency of the $t$-th communication round\\
				\hline
				$\Gamma_m$&Participation rate of the $m$-th gateway and its associated devices\\
				\hline
	\end{tabular}}}
\end{table}

\section{Preliminaries}\label{sec:Preliminaries}
\subsection{Federated Learning}
FL enables local model training across distributed devices, and global model aggregation at a centralized server. Considering an FL network of $N$ devices collaborating to train an ML model over their respective local datasets. The goal of FL is to find a set of model parameters $\boldsymbol{w}$ that minimizes the global loss function $F(\boldsymbol{w})$ on all the local datasets, i.e., $F(\boldsymbol{w})=\frac{\sum_n\vert\mathcal{D}_n\vert F_n(\boldsymbol{w})}{\sum_n\vert\mathcal{D}_n\vert}$, where $F_n(\boldsymbol{w})=f(\boldsymbol{w},\mathcal{D}_n)$ denotes the loss function on the local dataset $\mathcal{D}_n$. To keep the training data localized and private, each device in FL utilizes the gradient-descent method to minimize the local loss function $F_n(\boldsymbol{w})$ over its local dataset by iteratively moving in the negative direction of the gradient. To obtain the global model, they synchronously upload the trained local model parameters to the centralized server, which aggregates all the collected local model parameters and returns the result to each device to update the local model parameters. Compared with traditional ML, FL can collaboratively build a shared model without raw data exchange, which greatly reduces the communication overhead and promotes the privacy of localized data.
\subsection{Deep Neural Network Partition}
\subsubsection{Deep Neural Network}
A deep neural network (DNN) can be considered as stacked layers of neural networks, where raw data gets passed to the input layer and the output layer outputs the prediction result. Each hidden layer takes in the inputs, passes the weighted inputs into an activation function along with the biases, and forwards the outputs to the next layer.
\subsubsection{Forward and Backward Propagation}
According to the gradient-descent method, the backpropagation algorithm is utilized to calculate the gradient of the objective loss function with respect to the model parameters (i.e., neural network's weights and biases) by the chain rule\cite{DBLP:journals/tsp/BenvenutoP92}. Specifically, forward and backward propagation are executed in each iteration until the objective loss function converges. In the forward propagation stage, each hidden layer calculates the outputs by adding the biases to the weighted inputs and passing results into the activation function. The data flows from the first input layer to the last output layer to obtain the prediction result, where the output of each hidden layer serves as the input of the next one. In the backward propagation stage, error propagates in the opposite direction from the output layer to the input layer in order to compute the gradient of the loss function with regard to the model parameters. The error term in the output layer is calculated as the difference of actual and desired output, and the error term in each hidden layer is calculated as the weighted sum of the errors from the next layer. Based on the error passed from the next layer, each layer computes the gradient to update the model parameters and the error to be propagated to the previous layer.
\subsubsection{Deep Neural Network Partition}
In DNN partition mechanism, we set a partition point to divide the objective DNN into two continuous portions, and separately deploy the bottom and top layers of the DNN at an end device and an edge server. To perform the forward and backward propagation, the end device first transmits the labels of training dataset to the edge server. Then, the output of the last layer in the device-side DNN is transmitted to the edge server during the forward propagation stage, while the error term of first layer in the server-side DNN is transmitted to the end device during the backward propagation stage. Based on DNN partition mechanism, the local model training of top DNN portion is offloaded to the edge server, thereby greatly reducing the computational burden on the resource-constrained device side. Upon completing the DNN model training, we perform the model combination, i.e., combining the bottom and top layers of the trained DNN model to obtain the complete DNN model. The decision of DNN partition point mainly depends on three aspects as follows: (a) computational resources (i.e., processing power, memory capacity, etc.) of edge server. The computational resources required by the training of the offloaded DNN layers cannot exceed the computational resources of the edge server; (b) communication overhead. In the training process, DNNs can be partitioned in pooling layers to reduce the data size of forward outputs and errors transmitted between the end device and edge server; (c) privacy concern. Deeper DNN partition points can help mitigating the potential threats of privacy leakage.
\begin{figure*}[!t]
	\centering
	\includegraphics[width=6.1in]{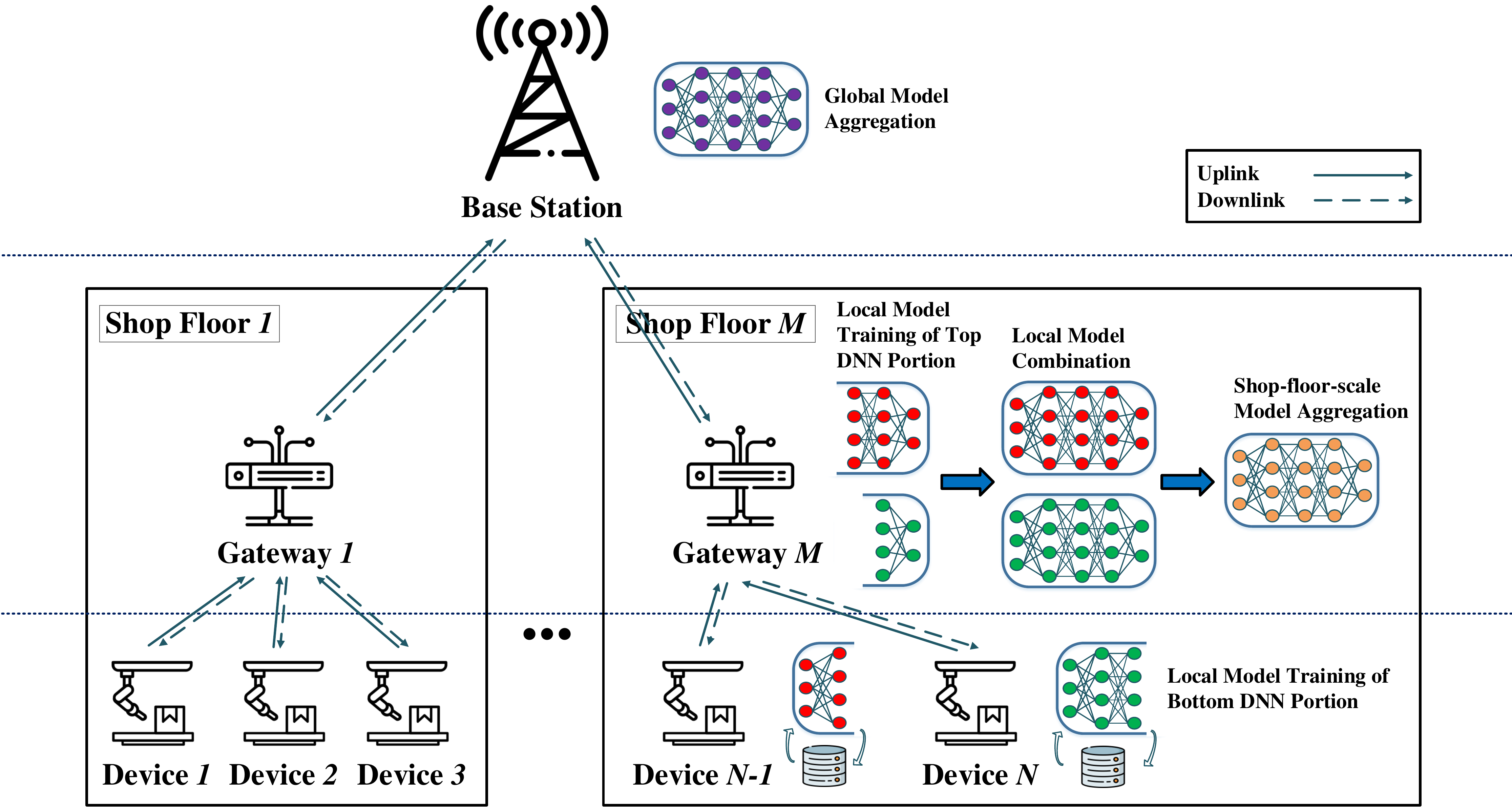}\vspace{6pt}
	\caption{System model of a two-tier communication-computation efficient FL-enabled IIoT framework.}
	\label{fig: system model}
\end{figure*}
\section{System Model}\label{sec:System Model}
\subsection{System Overview}
As shown in Fig.\ref{fig: system model}, we consider an FL-enabled IIoT network with $M$ shop floors, wherein each shop floor employs a single edge gateway and a group of end devices. In every shop floor, each device monitors the manufacturing process, collects local datasets, and performs local model training. Due to limited computation and memory resources at the device side, each device trains bottom layers of the objective DNN locally, and the training of top layers is offloaded to the edge gateway in the same shop floor. Then, the gateway collects the local models, combines the bottom and top DNN portions, and performs the shop-floor-scale model aggregation. The base station (BS) collects the aggregated shop-floor-scale models, and performs the global model aggregation to obtain a shared model.
\subsubsection{End devices}
Let $\mathcal{N}=\{1,\ldots,N\}$ and $\mathcal{M}=\{1,\ldots,M\}$ denote the index sets of the devices and gateways, respectively. Define an $N\times M$ deployment matrix as $\boldsymbol{a}$ with entry $a_{n,m}\in\{0,1\}$, $n\in\mathcal{N}$ and $m\in\mathcal{M}$. If $a_{n,m}=1$, the $n$-th device is deployed with the $m$-th gateway in the $m$-th shop floor. As such, the deployment matrix satisfies $\sum_{n\in\mathcal{N}}a_{n,m}=1$, $\forall m\in\mathcal{M}$. Each group of devices can only communicate with the gateway in the same shop floor, and we describe the group of devices as the associated devices with the gateway in the same shop floor. Each device holds a local dataset $\mathcal{D}_n=\{\boldsymbol{x}_{n,i}\in\mathbb{R}^d, y_{n,i}\in\mathbb{R}\}_{i=1}^{D_n}$ with $D_n=\vert\mathcal{D}_n\vert$ data points, where $\boldsymbol{x}_{n,i}$ and $y_{n,i}$ are the feature vector and label for the $i$-th data point at the $n$-th device. Let $\mathcal{L}=\{1,\ldots,L\}$ denote the index set of the DNN layers. For local model training, the bottom $l_n(t)$ layers of the objective DNN are trained locally at the $n$-th device in the $t$-th communication round, while the training of top $L-l_n(t)$ layers are offloaded to the associated gateway.

\subsubsection{Edge gateways}
The roles of each gateway include local model training of top DNN portion, device-level model combination, shop-floor-scale model aggregation, and model transmission. First, each gateway performs the forward and backward propagation for the top layers of the objective DNNs offloaded from the associated devices. Second, each gateway collects the bottom layers of the DNNs from the associated devices, combines the bottom and top layers of the trained DNNs, aggregates the combined local models, and transmits the aggregated shop-floor-scale model to the BS. Note that only a part of the shop floors can be selected to participate in FL in each communication round.
\subsubsection{Base station}
Let $\mathcal{J}=\{1,\ldots,J\}$ denote the index set of the available channels, $\mathcal{T}=\{1,\cdots,T\}$ denote the index set of communication rounds. Orthogonal frequency-division multiplexing (OFDM) is adopted to transmit the shop-floor-scale model parameters from the gateways to the BS in parallel. In each communication round, $J$ selected gateways can communicate with the BS through the assigned channels. The BS equipped with a cloud server has two functions: (a) aggregating the model parameters received from the selected edge gateways; (b) sending back the global model parameters and the scheduling policy information (e.g., channel assignment) to the gateways.

Consider that FL operation in each communication round is synchronous. As Fig.\ref{fig: system model} illustrated, the FL in the $t$-th communication round operates in the following steps:
\begin{enumerate}
	\item At the beginning of the $t$-th communication round, the BS selects $J$ gateways according to scheduling policy, and broadcasts the global model parameters $\boldsymbol{W}^t$ to the selected gateways. Define the channel assignment matrix as $\boldsymbol{I}(t)$ with entry $I_{m,j}(t)\in\{0,1\}$, $m\in\mathcal{M}$ and $j\in\mathcal{J}$. If $I_{m,j}(t)=1$, the $m$-th gateway is assigned to the $j$-th channel in the $t$-th communication round. Note that the channel assignment matrix satisfies $\sum_{m\in\mathcal{M}}I_{m,j}(t)=1$ and $\sum_{j\in\mathcal{J}}I_{m,j}(t)\leq1$, $\forall t\in\mathcal{T}$. Then, each selected gateway broadcasts the global model parameters $\boldsymbol{W}^t$ to its associated devices.
	\item Upon receiving $\boldsymbol{W}^t$, each training device and the associated gateway collaboratively perform the forward and backward propagation by DNN partition mechanism to update the local model parameters. Let $\tilde{\boldsymbol{w}}_n^{0,t}=\boldsymbol{W}^t$ denote the initial local model parameters of the $n$-th device in the $t$-th communication round. For each training device and the associated gateway, local model parameters are updated according to the gradient-descent update rule with respect to the local loss function over a total of $K$ iterations. The update rule in the $k$-th iteration is $\tilde{\boldsymbol{w}}_n^{k,t}=\tilde{\boldsymbol{w}}_n^{k-1,t}-\beta\nabla \tilde{F}_n(\tilde{\boldsymbol{w}}_n^{k-1,t})$, where $\tilde{\boldsymbol{w}}_n^{k,t}$ denotes the local model parameters of the $n$-th device in the $k$-th iteration and the $t$-th communication round, $\beta>0$ is the step size, $\nabla\tilde{F}_n(\tilde{\boldsymbol{w}}_n^{k-1,t})=\nabla f_n(\tilde{\boldsymbol{w}}_n^{k-1,t},\tilde{\mathcal{D}}_n)$ is the stochastic gradient of local loss function, and $\tilde{\mathcal{D}}_n$ is a batch of the local dataset $\mathcal{D}_n$ with $\tilde{D}_n=\vert\tilde{\mathcal{D}}_n\vert$ sample points.
	\item Upon completing the local model training, each training device transmits the bottom layers of the DNN to the associated gateway. Then, the selected gateways combine the bottom and top layers of the trained DNNs, aggregate the combined local model parameters according to the federated averaging (FedAvg) algorithm\cite{DBLP:journals/tcom/YangLQP20}, i.e., $\hat{\boldsymbol{w}}_m^t=\frac{\sum_{n\in\mathcal{N}}a_{n,m}\tilde{D}_n\tilde{\boldsymbol{w}}_n^{K,t}}{\sum_{n\in\mathcal{N}}a_{n,m}\tilde{D}_n}$, and transmit the aggregated shop-floor-scale model parameters $\hat{\boldsymbol{w}}_m^t$ to the BS. With the received model parameters uploaded from the selected gateways, the BS updates the global model parameters by performing global aggregation according to FedAvg, i.e., $\boldsymbol{W}^{t+1}=\frac{\sum_{m\in\mathcal{M}}\sum_{j\in\mathcal{J}}I_{m,j}(t)D_m\hat{\boldsymbol{w}}_m^t}{\sum_{m\in\mathcal{M}}\sum_{j\in\mathcal{J}}I_{m,j}(t)D_m}$, where $D_m=\sum_{n\in\mathcal{N}}a_{n,m}\tilde{D}_n$.
\end{enumerate}

\subsection{Computation Model}
Before delving into the computation model, we first define the following notations for the hyper-parameters and tensor shapes in the forward and backward propagation. Let $B_s$ and $S_f$ denote the batch size and the precision format of the data type, respectively. For the convolution layer and pooling layer, $H_o$, $W_o$ and $C_o$ are output height, width, and channel, respectively; $H_i$, $W_i$ and $C_i$ are input height, width, and channel; $H_f$ and $W_f$ are the filter's height and width. For the fully connected layer, $S_i$ and $S_o$ are the input and output sizes. To calculate the memory usage and FLOPs for the bottom and top DNN portions trained at the device and gateway side, we list the main layer-level memory usage and FLOPs in Table \ref{table1} according to the backpropagation algorithm\cite{DBLP:conf/iiswc/SiuSMM18,DBLP:conf/bigdataconf/JustusBBM18}.
\begin{table*}[!t]
	\caption{Layer-level memory usage and FLOPs in DNN forward and backward propagation operations \label{table1}}
	\centering
	\renewcommand{\arraystretch}{1.2}
	\begin{tabular}{!{\vrule width 1pt}c|c|c|c|c!{\vrule width 1pt}}
		\noalign{\hrule height 1pt}
		\multirow{2}*{\textbf{Layer Category}}& \multicolumn{2}{c|}{\textbf{Memory Usage}} & \multicolumn{2}{c!{\vrule width 1pt}}{\textbf{Floating-point Operation}}\\\cline{2-5}
		~ &Tensor Category &Tensor Size & Operator Category& FLOPs\\
		\noalign{\hrule height 1pt}
		\multirow{4}*{Convolution} & Weight & $S_fC_iH_f W_fC_o$ & Forward Propagation &$2B_s C_iH_fW_fC_o H_o W_o$\\\cline{2-5}
		~ &Forward Outout &$S_f B_s C_o H_o W_o$ &  \multirow{2}*{Error Calculation} & \multirow{2}*{\tabincell{c}{$2 B_s (2 W_f+W_f W_o-2)$\\$\times(2 H_f+H_f H_o-2)$}}\\\cline{2-3}
		~ &Backward Error &$S_f B_s C_i H_i W_i$ & ~ & ~\\\cline{2-5}
		~ &Gradient &$S_fC_iH_fW_fC_o$ &Gradient Calculation & $2 B_s C_i H_f W_f C_o H_o W_o$\\
		\noalign{\hrule height 1pt}
		\multirow{2}*{Pooling} & Forward Outout &$S_f B_s C_o H_o W_o$ & Forward Propagation &$B_s C_iH_iW_i$\\\cline{2-5}
		~ &Backward Error &$S_f B_s C_i H_i W_i$ & Error Calculation & $B_s C_iH_iW_i$\\
		\noalign{\hrule height 1pt}
		\multirow{4}*{Fully Connected} & Weight & $S_iS_o$ & Forward Propagation &$2 B_s S_i S_o$\\\cline{2-5}
		~ &Forward Outout &$B_s S_o$ & Error Calculation & $2 B_s S_i S_o$\\\cline{2-5}
		~ &Backward Error &$B_s S_i$ & \multirow{2}*{Gradient Calculation} & \multirow{2}*{$B_sS_iS_o$}\\\cline{2-3}
		~ &Gradient &$S_iS_o$ &~ & ~\\
		\noalign{\hrule height 1pt}
	\end{tabular}
\end{table*}

Let $o_l$ and $o'_l$ denote the FLOPs of the forward and backward propagation for each sample point in the $l$-th layer, respectively. As such, the local model training time of the $m$-th gateway and its associated devices in the $t$-th communication round is represented as
\begin{alignat}{1}
	\tau_m^\text{tra}(t)&=\sum_{j\in\mathcal{J}}I_{m,j}(t)\max_{n\in\mathcal{N}}\left\{a_{m,n}K\tilde{D}_n\left(\frac{\sum_{l=1}^{l_n(t)}(o_l+o'_l)}{\phi_n^\text{D}f_n^\text{D}}\right.\right.\nonumber\\&\left.\left.+\frac{\sum_{l=l_n(t)+1}^{L}(o_l+o'_l)}{\phi_m^\text{G}f_{m,n}^\text{G}(t)}\right)\right\},
\end{alignat}where $\phi_n^\text{D}$ and $\phi_m^\text{G}$ are the FLOPs per clock cycle of the $n$-th device and the $m$-th gateway, $f_n^\text{D}$ is the computation frequency of the $n$-th device for local model training, and $f_{m,n}^\text{G}(t)$ is the computation frequency of the $m$-th gateway assigned to the $n$-th device's local model training. Note that $f_{m,n}^\text{G}(t)$ is limited by the total computation frequency of the $m$-th gateway, i.e., $\sum_{n\in\mathcal{N}}a_{m,n}f_{m,n}^\text{G}(t)\leq f_m^\text{G,max}$. The energy consumption of the $n$-th device for local model training in the $t$-th communication round can be expressed as\cite{9542866}
\begin{alignat}{1}
	e_n^\text{tra,D}(t)=&\sum_{j\in\mathcal{J}}\sum_{m\in\mathcal{M}}I_{m,j}(t)a_{m,n}K\tilde{D}_n{v_n^\text{D}}/{\phi_n^\text{D}}\left({\sum}_{l=1}^{l_n(t)}(o_l\right.\nonumber\\&+o'_l)\bigg)\left(f_n^\text{D}\right)^2,
\end{alignat}where $v_n^\text{D}$ is the effective switched capacitance. Moreover, the energy consumption of the $m$-th gateway for the local model training offloaded from its associated devices in the $t$-th communication round is given by
\begin{alignat}{1}
	e_m^\text{tra,G}(t)=&\sum_{j\in\mathcal{J}}\sum_{n\in\mathcal{N}}I_{m,j}(t)a_{m,n}K\tilde{D}_nv_m^\text{G}/{\phi_m^\text{G}}\left({\sum}_{l=l_n(t)+1}^L\right.\nonumber\\&(o_l+o'_l)\bigg)\left(f_{m,n}^\text{G}(t)\right)^2.
\end{alignat}

For the $n$-th training device with the training dataset $\tilde{\mathcal{D}}_n$, let $g_{n,l}$ denote the memory usage of the $l$-th layer for storing the model parameters and intermediate data in the forward and backward propagation. The total memory usage for the bottom and top layers of the objective DNN, which are trained at the device and gateway side, are given by\begin{alignat}{1} G_n^\text{D}(t)=\sum_{j\in\mathcal{J}}\sum_{m\in\mathcal{M}}\sum_{l=1}^{l_n(t)}I_{m,j}(t)a_{m,n}g_{n,l},\end{alignat} and \begin{alignat}{1}G_m^\text{G}(t)=\sum_{j\in\mathcal{J}}\sum_{n\in\mathcal{N}}\sum_{l=l_n(t)+1}^{L}I_{m,j}(t)a_{m,n}g_{n,l}.\end{alignat} Let $G_n^\text{D,max}$ and $G_m^\text{G,max}$ denote the memory size of the $n$-th device and the $m$-th gateway, respectively. We can note that $0\leq G_n^\text{D}(t)\leq G_n^\text{D,max}$ and $0\leq G_m^\text{G}(t)\leq G_m^\text{G,max}$, since the memory usage cannot exceed the memory size of the equipment. In addition, the local model training cannot be fully offloaded due to the limited memory and energy resources at the edge gateway side.

\subsection{Communication Model}
At the beginning of the $t$-th communication round, the BS broadcasts the global model parameters to the selected gateways over wireless channels. Assume that the wireless channels are IID block fading. The channel remains static in each communication round but varies among different communication rounds. In our communication model, the downlink channel power gain from the BS to the $m$-th gateway via the $j$-th channel is modeled as $h_{m,j}^\text{d}(t)=h_0\rho_{m,j}^\text{d}(t)(d_0/d_m)^{\nu}$, where $h_0$ is the path loss constant, $\rho_{m,j}^\text{d}(t)$ is the small-scale fading channel power gain from the BS to the $m$-th gateway via the $j$-th channel in the $t$-th communication round, $d_m$ is the distance from the BS to the $m$-th gateway, $d_0$ is the reference distance, and $\nu$ is the large-scale path loss factor, respectively. Thus, the global model transmission time from the BS to the $m$-th gateway in the $t$-th communication round can be represented as\begin{equation} \tau_m^\text{down}(t)=\sum_{j\in\mathcal{J}}\frac{I_{m,j}(t)\gamma}{B^\text{d}\log_2\left(1+\frac{P^Bh_{m,j}^\text{d}(t)}{B^\text{d}N_0+i^\text{d}_{m,j}(t)}\right)},\end{equation}where $\gamma$ is DNN model size, $B^\text{d}$ is the bandwidth of the downlink channel, $P^B$ is the transmit power of the BS, $N_0$ is the noise power spectral density, and $i^\text{d}_{m,j}(t)$ is the co-channel interference caused by radio communication services in other areas, respectively.

After downloading the global model parameters from the BS, each selected gateway broadcasts the global model parameters to its associated devices. Due to the short-distance wireless technology, we consider that the transmission time between the gateways and the associated devices is negligible compared with the overall FL training delay\cite{DBLP:journals/vlsisp/WuCL18,9678373,9066337}. After completing the local model training, each training device transmits the bottom layers of the DNN to its associated gateway, and the selected gateways perform the device-level model combination and forward the aggregated shop-floor-scale model parameters to the BS over wireless links. Similarly, the model transmission time from the $m$-th gateway to the BS in the $t$-th communication round is
\begin{equation}
	\tau_m^\text{up}(t)=\sum_{j\in\mathcal{J}}\frac{I_{m,j}(t)\gamma}{B^\text{u}\log_2\left(1+\frac{P_m(t)h^u_{m,j}(t)}{B^\text{u}N_0+i^u_{m,j}(t)}\right)},
\end{equation}where $B^\text{u}$ represents the bandwidth of the uplink channel, $P_m(t)$ denotes the transmit power of the $m$-th gateway, $i^u_{m,j}(t)$ is the co-channel interference, $h^u_{m,j}(t)=h_0\rho_{m,j}^\text{u}(t)(d_0/d_m)^{\nu}$ is the uplink channel power gain from the $m$-th gateway to the BS, and $\rho_{m,j}^\text{u}(t)$ is the small-scale fading channel power gain, respectively. The energy consumption of the $m$-th gateway for transmitting the aggregated shop-floor-scale model parameters in the $t$-th communication round is
\begin{equation}
	e_m^\text{up}(t)=\sum_{j\in\mathcal{J}}\frac{P_m(t)I_{m,j}(t)\gamma}{B^\text{u}\log_2\left(1+\frac{P_m(t)h^u_{m,j}(t)}{B^\text{u}N_0+i^u_{m,j}(t)}\right)}.
\end{equation}

In addition, the energy harvesting (EH) components equipped at devices and gateways harvest renewable energy from the nature for local model training and transmission. We formulate the EH process as successive energy packet arrivals. Let $E_n^\text{D}(t)$ and $E_m^\text{G}(t)$ denote the energy arrival at the $n$-th device and the $m$-th gateway in the $t$-th communication round. Consider that $E_n^\text{D}(t)$ and $E_m^\text{G}(t)$ are modeled as IID stochastic processes, i.e., $E_n^\text{D}(t)$ and $E_m^\text{G}(t)$ are uniformly distributed within $[0, E_n^\text{D,max}]$ and $[0, E_m^\text{G,max}]$, respectively. Note that the total energy consumption of the $m$-th gateway in the $t$-th communication round can be represented as \begin{equation}e_m^\text{G}(t)=e_m^\text{tra,G}(t)+e_m^\text{up}(t).\end{equation}As such, it can be derived that $0\leq e_n^\text{tra,D}(t)\leq E_n^\text{D}(t)$, and $0\leq e_m^\text{G}(t)\leq E_m^\text{G}(t)$, since the energy consumption cannot exceed the energy arrivals in each communication round.

\subsection{Problem Formulation}
According to the analysis above, the time consumption of each communication round mainly comes from three parts, i.e., global model downloading, local model training, and shop-floor-scale model uploading. Thus, the total delay of the $t$-th communication round is given by
\begin{equation}
	\tau(t)=\max_{m\in\mathcal{M}}\left\{\tau_m^\text{tra}(t)+\tau_m^\text{up}(t)+\tau_m^\text{down}(t)\right\}.
\end{equation} To obtain a communication-computation efficient FL framework, we develop a dynamic device selection and resource scheduling protocol to minimize average delay under the energy consumption and memory usage constraints. Let $\boldsymbol{X}(t) = [\boldsymbol{I}(t), \boldsymbol{l}(t), \boldsymbol{P}(t), \boldsymbol{f}^\text{G}(t)]$. In this context, we formulate a stochastic optimization problem as\begin{alignat}{1}\label{P0}
&\quad\quad\quad\quad\textbf{P0}:\;\min_{\boldsymbol{X}(t)}\; \frac{1}{T}{\sum}_{t=1}^T\tau(t)\\
\!\!\!\!\!\!\!\!\!\!\!\!\!\!\!\!\!\!\!\text{s.t.}\quad
&\textbf{C1}: I_{m,j}(t)\in \{0,1\}, \forall m\in \mathcal{M}, j\in \mathcal{J}, t\in \mathcal{T},\nonumber\\
&\textbf{C2}: \sum_{j\in\mathcal{J}}I_{m,j}(t)\leq 1, \forall m\in \mathcal{M}, t\in \mathcal{T},\nonumber\\
&\textbf{C3}: \sum_{m\in\mathcal{M}}I_{m,j}(t)=1, \forall j\in \mathcal{J}, t\in \mathcal{T},\nonumber\\
&\textbf{C4}: 0\leq P_m(t)\leq P_m^\text{\rm{max}}, \forall m\in \mathcal{M}, t\in \mathcal{T},\nonumber\\
&\textbf{C5}: 0\leq l_n(t)\leq L, \forall n\in \mathcal{N}, t\in \mathcal{T},\nonumber\\
&\textbf{C6}: f_m^\text{G,min}\leq \sum_{n\in\mathcal{N}}a_{m,n}f_{m,n}^\text{G}(t)\leq f_m^\text{G,max},\forall m\in \mathcal{M}, t\in \mathcal{T},\nonumber\\
&\textbf{C7}: 0\leq G_n^\text{D}(t)\leq G_n^\text{D,max}, \forall n\in \mathcal{N},t\in \mathcal{T},\nonumber\\
&\textbf{C8}: 0\leq G_m^\text{G}(t)\leq G_m^\text{G,max}, \forall m\in \mathcal{M},t\in \mathcal{T},\nonumber\\
&\textbf{C9}: 0\leq e_n^\text{tra,D}(t)\leq E_n^\text{D}(t), \forall n\in \mathcal{N},t\in \mathcal{T},\nonumber\\
&\textbf{C10}: 0\leq e_m^\text{G}(t)\leq E_m^\text{G}(t), \forall m\in \mathcal{M},t\in \mathcal{T},\nonumber\\
&\textbf{C11}: \frac{1}{T}\sum_{t=1}^T\mathbbm{1}_m^t\ge\Gamma_m, \forall m\in \mathcal{M},\nonumber
\end{alignat}where $\mathbbm{1}_m^t=\sum_{j\in\mathcal{J}}I_{m,j}(t)$ indicates whether the $m$-th gateway is selected to participate in the local model training in the $t$-th communication round. That is, if $\mathbbm{1}_m^t=1$, the $m$-th gateway and associated devices are selected to train the local model in the $t$-th communication round. $\Gamma_m$ is the participation rate of the $m$-th gateway and its associated devices derived in the following section. The ranges of the variables $\boldsymbol{I}(t)$, $\boldsymbol{l}(t)$, $\boldsymbol{P}(t)$ and $ \boldsymbol{f}^\text{G}(t)$ are constrained by $\textbf{C1}\sim\textbf{C6}$, respectively. $\textbf{C7}\sim\textbf{C10}$ are the memory usage and energy consumption constraints for devices and gateways in each communication round, respectively. Furthermore, the long-term constraint $\textbf{C11}$ is adopted to optimize the FL performance by guaranteeing the participation rate for each gateway and the associated devices. Overall, the goal of $\textbf{P0}$ is to jointly optimize communication and computation resources under memory usage, energy consumption and participation rate constraints.

\section{Device-specific Participate Rate}\label{sec:Device-specific measurement of the participation rate}
In this section, we derive a model divergence bound to measure the learning performance of each gateway and the associated devices' local model training. As such, the participation rate of each gateway and the associated devices can be determined based on the derived divergence bound. Our analysis of the model divergence bound focuses on three parts, i.e., data distribution, training dataset size, and the number of local epochs.

Before the analysis, two auxiliary notations are introduced. We use $\boldsymbol{w}_n^{k,t}$ to denote the set of local model parameters that follows a full gradient descent, i.e., $\boldsymbol{w}_n^{k+1,t}=\boldsymbol{w}_n^{k,t}-\beta\nabla f(\boldsymbol{w}_n^{k,t},\mathcal{D}_n)$, and $\boldsymbol{v}^{k,t}$ to denote the set of local model parameters that follows a centralized gradient descent, i.e., $\boldsymbol{v}^{k+1,t}=\boldsymbol{v}^{k,t}-\beta\nabla f(\boldsymbol{v}^{k,t},\cup\mathcal{D}_n)$. Note that although the sets of model parameters $\tilde{\boldsymbol{w}}_n^{k,t}$, $\boldsymbol{w}_n^{k,t}$, and $\boldsymbol{v}^{k,t}$ follow different update rules, they are synchronized with $\tilde{\boldsymbol{w}}^{K,t-1}$ at the beginning of the $t$-th communication round, i.e., $\tilde{\boldsymbol{w}}^{0,t}=\boldsymbol{w}^{0,t}=\boldsymbol{v}^{0,t}=\tilde{\boldsymbol{w}}^{K,t-1}$. In addition, let $\tilde{\boldsymbol{w}}^{k,t}=\sum_n\frac{\sum_m \mathbbm{1}_m^ta_{m,n}\tilde{D}_n}{\sum_n\sum_m \mathbbm{1}_m^ta_{m,n}\tilde{D}_n}\tilde{\boldsymbol{w}}_n^{k,t}$ and $\boldsymbol{w}^{k,t}=\sum_{n\in\mathcal{N}}\frac{D_n}{\sum_{n\in\mathcal{N}}D_n}\boldsymbol{w}_n^{k,t}$ denote the weighted average of the sets of model parameters $\tilde{\boldsymbol{w}}_n^{k,t}$ and $\boldsymbol{w}_n^{k,t}$, respectively.

To facilitate the analysis, we make the following assumptions on the loss function to describe how the data is distributed at different devices.
\begin{assumption}\label{assumption3}
	For each data point $\{\boldsymbol{x}_i,y_i\}\in\mathcal{D}_n$, the gradient of the function $ f(\boldsymbol{w},\{\boldsymbol{x}_i,y_i\})$ has bounded variance, i.e., $\mathbb{E}{\Vert\nabla f(\boldsymbol{w},\{\boldsymbol{x}_i,y_i\})-\nabla f(\boldsymbol{w},\mathcal{D}_n)\})\Vert}\leq\sigma_n$.
\end{assumption}
\begin{assumption}\label{assumption4}
	For each device, the gradient of the local loss function $f(\boldsymbol{w},\mathcal{D}_n)$ and the global loss function $f(\boldsymbol{w},\cup\mathcal{D}_n)$ satisfy $\Vert\nabla f(\boldsymbol{w},\mathcal{D}_n)-\nabla f(\boldsymbol{w},\cup\mathcal{D}_n)\Vert\leq \delta_n$.
\end{assumption}

Based on \textbf{Assumption} \ref{assumption3} and \ref{assumption4}, we investigate model divergence $\left\Vert\hat{\boldsymbol{w}}_m^t\!-\!\boldsymbol{v}^{K,t}\right\Vert$ in \textbf{Theorem} \ref{theorem1}.
\begin{theorem}\label{theorem1}
	Assume that the local loss function $f(\boldsymbol{w},\mathcal{D}_n)$ is $L_n$-smooth. The divergence between $\hat{\boldsymbol{w}}_m^t$ and $\boldsymbol{v}^{K,t}$ in the $t$-th communication round can be written as\begin{alignat}{1}\label{theorem1_1}
		\left\Vert\hat{\boldsymbol{w}}_m^t-\boldsymbol{v}^{K,t}\right\Vert\leq\Phi_m\triangleq& \sum_{n\in\mathcal{N}}\frac{a_{m,n}\tilde{D}_n}{\sum_{n\in\mathcal{N}}a_{m,n}\tilde{D}_n}\left(\frac{\sigma_n}{L_n\sqrt{\tilde{D}_n}}\right.\nonumber\\&+\frac{\delta_n}{L_n}\Bigg)\left((\beta L_n+1)^k-1\right).
	\end{alignat}
\end{theorem}
\begin{IEEEproof}
	Please see Appendix A.
\end{IEEEproof}

\textbf{Theorem} \ref{theorem1} reveals the impact of data distribution on FL performance, wherein lower variances $\sigma_n$ and $\delta_n$ produce better training performance. That is, the gateway and associated devices are more helpful for FL training if the local data distribution better represents the overall data distribution. In addition, we find that larger training data size $\tilde{D}_n$ can lead to smaller divergence. Moreover, the divergence increases with the value of local epoch $K$, which follows the same trend with the standard FL framework\cite{DBLP:journals/jsac/WangTSLMHC19}.

According to the model divergence bound in (\ref{theorem1_1}), we derive the proportion of the $m$-th gateway and its associated devices’ participation rate over the total participation rate as $\frac{1/\Phi_m}{\sum_{m\in\mathcal{M}}1/\Phi_m}$. Recall that we select $J$ gateways and the associated devices to participate in the local model training in each communication round. In this context, the total participation rate of all gateways and the associated devices is $J$. As such, the participation rate of the $m$-th gateway and its associated devices is determined by\cite{DBLP:journals/jsac/LyuR00LD19,DBLP:journals/twc/XiaQGWYZ20,DBLP:journals/tpds/HuangLWHLZ21}\begin{equation}\label{12}
	\Gamma_m=\min\left\{J\frac{1/\Phi_m}{\sum_{m\in\mathcal{M}}1/\Phi_m},1\right\}.
\end{equation}Note that the participation rate of each gateway and its associated devices cannot exceed $1$.

This participation rate $\Gamma_m$ derived by the divergence bound of the $m$-th gateway and associated devices is introduced to show how many communication rounds that the $m$-th gateway should participate in the whole FL process. Based on the derived participation rate $\Gamma_m$, we can optimize the communication and energy resources while guaranteeing FL training performance by adopting a device-specific participation rate constraint. Superior to the general fairness guarantee (e.g. Round Robin), the participation rate constraint can not only save the slow devices from being excluded from FL training process, but also involve important devices with better data distribution in more communication rounds on the track of low latency by setting a larger participation rate for the important devices.

\begin{algorithm}[!t]
		\caption{Dynamic device scheduling and resource allocation algorithm}\label{alg::Dynamic Resource Allocation and Client Scheduling Algorithm}
		Initialize:	Virtual queue length $\boldsymbol{Q}(t)=0$;\\
		\For{$t=1,2,...,T$}{
			\algorithmicrequire{ Virtual queue length and channel state at the beginning of the $t$-th communication round};\\
			\algorithmicensure{ $\boldsymbol{X}(t)=$ [$\boldsymbol{I}(t)$, $\boldsymbol{l}(t)$, $\boldsymbol{P}(t)$, $\boldsymbol{f}^\text{G}(t)$]};\\
			\DontPrintSemicolon
			\SetKwBlock{DoParallel}{do in parallel}{end}
			\DoParallel{
				Optimize DNN partirion point $\boldsymbol{l}(t)$, computation frequency $\boldsymbol{f}^\text{G}(t)$ and transmit power $\boldsymbol{P}(t)$ by solving (\ref{P3_1_1}), (\ref{P3_1_2}), and (\ref{P3_1_3}) with block coordinate descent method, and compute $\Lambda_{m,j}(t)$ according to (\ref{Lambda});
			}
		    Given the optimized auxiliary variable $\Lambda_{m,j}(t)$, find the channel assignment policy $\boldsymbol{I}(t)$ by solving (\ref{P3_2_1}) with Hungarian method;\\
			Update $\boldsymbol{Q}(t)$ according to (\ref{vitual queues});\\
			\algorithmicreturn{ $\boldsymbol{X}(t)=$ [$\boldsymbol{I}(t)$, $\boldsymbol{l}(t)$, $\boldsymbol{P}(t)$, $\boldsymbol{f}^\text{G}(t)$]}
		}
\end{algorithm}
\section{Dynamic Device Scheduling and Resource Allocation Algorithm}\label{sec:Dynamic channel assignment and resource allocation algorithm}
In this section, we propose a dynamic device scheduling and resource allocation (DDSRA) algorithm to solve the stochastic optimization problem \textbf{P0}, which is shown in \textbf{Algorithm} \ref{alg::Dynamic Resource Allocation and Client Scheduling Algorithm}. The proposed DDSRA as a centralized scheduling algorithm is performed by the BS. Compared with the existing DNN partition approaches using a predefined DNN partition point for all devices during the FL training process \cite{DBLP:conf/mlicom/ZhangWZ019,DBLP:conf/ics/DemirciF21,DBLP:journals/monet/ZhangZWC20}, the proposed DDSRA algorithm dynamically optimizes DNN partition point, channel assignment, transmit power, and computation frequency with time-varying channels and stochastic energy arrivals.
\subsection{Problem Transformation}
Based on the Lyapunov optimization method\cite{DBLP:series/synthesis/2010Neely}, we first transform the original problem \textbf{P0} into \textbf{P1} by converting the time-average inequality constraint \textbf{C11} to the queue stability constraint \textbf{C11'}. To this end, we define the virtual queue $Q_m(t)$ for each gateway updated by\begin{equation}\label{vitual queues}
	Q_m(t+1)\triangleq\max\left\{Q_m(t)-\mathbbm{1}_m^t+\Gamma_m,0\right\}.
\end{equation}By replacing the long-term participation rate constraint $\textbf{C11}$ with mean rate stability constraint of $Q_m(t)$, the original problem \textbf{P0} can be written as\begin{alignat}{1}\label{P1}
	\textbf{P1}:\;&\max_{\boldsymbol{X}(t)}\;\frac{1}{T}{\sum}_{t=1}^T\tau(t)\\
	\text{s.t.}\quad
	&\textbf{C1}\sim\textbf{C10},\;\textbf{C11'}:\lim\limits_{t\to\infty}\frac{\mathbb{E}\{\vert Q_m(t)\vert\}}{t}=0, \forall m\in\mathcal{M}.\nonumber
\end{alignat}

To solve \textbf{P1}, we next transform the long-term stochastic problem \textbf{P1} into the static problem \textbf{P2} in each communication round by means of characterizing the Lyapunov drift-plus-penalty function\cite{DBLP:series/synthesis/2010Neely}.
\begin{definition}
	Given $V> 0$, the Lyapunov drift-plus-penalty function is defined as\begin{equation}\label{drift-plus-penalty function}
		\Delta_V(t)\triangleq V\tau(t)+\Delta \Xi(t),
	\end{equation}where $\Delta \Xi(t)\triangleq\mathbb{E}\{\Xi(t+1)-\Xi(t)\vert\boldsymbol{Q}(t)\}$ is the conditional Lyapunov drift, and $\Xi(t)\triangleq\frac{1}{2}\sum_{m\in\mathcal{M}}Q_m(t)^2$ is the Lyapunov function.
\end{definition}

Minimizing $\Delta \Xi(t)$ stabilizes the virtual queues $\boldsymbol{Q}(t)$ and encourages the virtual queues to meet the mean rate stability constraint \textbf{C11'}\cite{DBLP:journals/corr/abs-2105-14708}. As such, minimizing the Lyapunov drift-plus-penalty function can concurrently minimize the FL delay and satisfy the long-term participation rate constraint $\textbf{C11}$, where $V$ is a control parameter to tune the trade-off between latency minimization and the degree of which the long-term participation rate constraint is satisfied.
\begin{lemma}\label{bound of Deltat}
	Given the virtual queue lengths $\boldsymbol{Q}(t)$, $\Delta \Xi(t)$ is upper bounded by $\Delta \Xi(t)\leq H+{\sum}_{m\in \mathcal{M}}$ $\mathbb{E}\left\{\left.Q_m(t)(\Gamma_m-\mathbbm{1}_m^t)\right\vert\boldsymbol{Q}(t)\right\}$, where $H=\frac{1}{2}\sum_{m\in \mathcal{M}}(\Gamma_m+1)$.
\end{lemma}
\begin{IEEEproof}
	Please see Appendix B.
\end{IEEEproof}

Thus, the DDSRA algorithm is proposed to minimize the Lyapunov drift-plus-penalty function $\Delta_V(t)$ in (\ref{drift-plus-penalty function}) in each communication round, i.e.,
\begin{alignat}{1}\label{P2}
	\textbf{P2}:\;&\min_{\boldsymbol{X}(t)}\;V\tau(t)-{\sum}_{m\in \mathcal{M}}{\sum}_{j\in\mathcal{J}}Q_m(t)I_{m,j}(t)\\
	\text{s.t.}\quad
	&\textbf{C1}\sim\textbf{C10}.\nonumber
\end{alignat}

\subsection{Optimal Solution of \textbf{P2}}
To solve \textbf{P2}, we first introduce an $M\times J$ matrix $\boldsymbol{\Lambda}(t)$ of auxiliary variables\begin{alignat}{1}\label{Lambda}
\Lambda_{m,j}(t)=&\max_{n\in\mathcal{N}_m}\left\{\left(\frac{\sum_{l=1}^{l_n(t)}(o_l+o'_l)}{\phi_n^\text{D}f_n^\text{D}}+\frac{\sum_{l=l_n(t)+1}^{L}(o_l+o'_l)}{\phi_m^\text{G}f_{m,n}^\text{G}(t)}\right.\right.\nonumber\\&\;K\tilde{D}_n\Bigg)\Bigg\}+{\gamma}/{B^\text{d}/\log_2\left(1+\frac{P^Bh_{m,j}^\text{d}(t)}{B^\text{d}N_0+i^\text{d}_{m,j}(t)}\right)}\nonumber\\&+{\gamma}/{B^\text{u}/\log_2\left(1+\frac{P_m(t)h^u_{m,j}(t)}{BN_0+i^u_{m,j}(t)}\right)}.
\end{alignat}Note that $\Lambda_{m,j}(t)$ represents the total delay for the $m$-th gateway if it is assigned to the $j$-th channel in the $t$-th communication round, and $\mathcal{N}_m\subset\mathcal{N}$ denotes the index set of the devices associated with the $m$-th gateway. As such, \textbf{P2} can be rewritten as
\begin{alignat}{1}\label{P3}
	\textbf{P3}:\;&\min_{\boldsymbol{X}(t)}\;V\max_{m\in\mathcal{M}}\left\{\sum_{j\in\mathcal{J}}I_{m,j}(t)\Lambda_{m,j}(t)\right\}-\!\!\sum_{m\in \mathcal{M}}\sum_{j\in\mathcal{J}}Q_m(t)\nonumber\\&\quad\quad\; I_{m,j}(t)\\
	\text{s.t.}\quad
	&\textbf{C1}\sim\textbf{C10}.\nonumber
\end{alignat}
By exploiting the independence between $\boldsymbol{I}(t)$ and $\boldsymbol{\Lambda}(t)$ in the objective function of \textbf{P3}, we decouple the joint optimization problem into the following sub-problems.
\subsubsection{Optimal auxiliary variable}
Since $\boldsymbol{\Lambda}(t)$ is independent of $\boldsymbol{I}(t)$ in \textbf{P3}, we can separately minimize $\Lambda_{m,j}(t)$ by optimizing the corresponding DNN partition point $l_n(t)$, transmitting power $P_m(t)$, and computation frequency $f_{m,n}^\text{G}(t)$ as\begin{alignat}{1}\label{P3_1}
	&\quad\quad\min_{l_n(t), P_m(t), f_{m,n}^\text{G}(t), \forall n\in\mathcal{N}_m}\;\Lambda_{m,j}(t)\\
	\text{s.t.}\quad &\textbf{C4}\sim\textbf{C6},\nonumber\\&\textbf{C7'}:{\sum}_{l=1}^{l_n(t)}g_{n,l}\leq G_n^\text{D,max},\forall n\in\mathcal{N}_m,\nonumber\\&\textbf{C8'}:\sum_{n\in\mathcal{N}_m}{\sum}_{l=l_n(t)+1}^{L}g_{n,l}\leq G_m^\text{G,max},\nonumber\\&\textbf{C9'}:\sum_{n\in\mathcal{N}_m}\!\!\!K\tilde{D}_n\frac{v_m^\text{G}}{\phi_m^\text{G}}{\left({\sum}_{l=l_n(t)+1}^L\!(o_l+o'_l)\!\right)}\!\left(f_{m,n}^\text{G}(t)\right)^2\nonumber\\&\quad\quad+\frac{\gamma P_m(t)}{B^\text{u}\log_2\left(1+\frac{P_m(t)Bh_{m,j}^\text{u}(t)}{B^\text{u}N_0+i^\text{u}_{m,j}(t)}\right)}\leq E_m^\text{G}(t),\nonumber\\&\textbf{C10'}:K\tilde{D}_n\frac{v_n^\text{D}}{\phi_n^\text{D}}\!\!\left({\sum_{l=1}^{l_n(t)}\!(o_l\!+\!o'_l)}\!\!\right)\!\!\!\left(f_n^\text{D}\right)^{\!2}\!\!\leq\! E_n^\text{D}(t),\forall n\in\mathcal{N}_m.\nonumber
\end{alignat}To solve this problem, we decompose (\ref{P3_1}) into three sub-problems in (\ref{P3_1_1}), (\ref{P3_1_2}) and (\ref{P3_1_3}). Given the remaining variables, each sub-problem is solved by the bisection method or successive convex optimization method\cite{DBLP:journals/tsp/FengP15a}. Thus, (\ref{P3_1}) can be optimized by the block coordinate descent method as shown in \textbf{Algorithm} \ref{alg::Dynamic Resource Allocation and Client Scheduling Algorithm}.

Given the optimized $P_m(t)$ and $f_{m,n}^\text{G}(t)$, we can rewrite (\ref{P3_1}) as\begin{alignat}{1}\label{P3_1_1}
	\min_{l_n(t),\forall n\in\mathcal{N}_m}\;&g_1(l_n(t))=\max_{n\in\mathcal{N}_m}\left\{K\tilde{D}_n\left(\frac{\sum_{l=1}^{l_n(t)}(o_l+o'_l)}{\phi_n^\text{D}f_n^\text{D}}\right.\right.\nonumber\\&\quad\quad\quad\;\;\left.\left.+\frac{\sum_{l=l_n(t)+1}^{L}(o_l+o'_l)}{\phi_m^\text{G}f_{m,n}^\text{G}(t)}\right)\right\}\\
	\text{s.t.}\quad &\textbf{C5}, \textbf{C7'}, \textbf{C8'}, \textbf{C9'}, \textbf{C10'}.\nonumber
\end{alignat}Note that the sub-problem in (\ref{P3_1_1}) is NP-hard. To circumvent this difficulty, a greedy solution with polynomial-time complexity is proposed by adopting the bisection method\cite{DBLP:journals/tie/HoFP20}. Let $g_1^\text{min}=\frac{K\min_{n\in\mathcal{N}_m}\{\tilde{D}_n\}\sum_{l=1}^{L}(o_l+o'_l)}{\max\{\max_{n\in\mathcal{N}_m}\{\phi_n^\text{D}f_n^\text{D}\},\max_{n\in\mathcal{N}_m}\{\phi_m^\text{G}f_{m,n}^\text{G}(t)\}\}}$ and $g_1^\text{max}=\frac{K\max_{n\in\mathcal{N}_m}\{\tilde{D}_n\}\sum_{l=1}^{L}(o_l+o'_l)}{\min\{\min_{n\in\mathcal{N}_m}\{\phi_n^\text{D}f_n^\text{D}\},\min_{n\in\mathcal{N}_m}\{\phi_m^\text{G}f_{m,n}^\text{G}(t)\}\}}$ denote the lower and upper bound of $g_1(l_n(t))$. Let $\eta$ be the mid point of the interval $(g_1^\text{min}, g_1^\text{max})$, i.e., $\eta=\frac{1}{2}(g_1^\text{min}+g_1^\text{max})$. In each iteration, we first compute the lower and upper bound of $l_n(t)$ according to constraints \textbf{C5}, \textbf{C7'}, \textbf{C9'} and $K\tilde{D}_n\left(\frac{\sum_{l=1}^{l_n(t)}(o_l+o'_l)}{\phi_n^\text{D}f_n^\text{D}}+\frac{\sum_{l=l_n(t)+1}^{L}(o_l+o'_l)}{\phi_m^\text{G}f_{m,n}^\text{G}(t)}\right)\leq \eta$, $\forall n\in\mathcal{N}_m$, i.e., $l_n^\text{min}\leq l_n(t)\leq l_n^\text{max}$. If constraints \textbf{C8'} and \textbf{C10'} hold when $l_n(t)=l_n^\text{min}$, we refine the upper bound of $g_1(l_n(t))$ as $\eta$. Otherwise, the lower bound of $g_1(l_n(t))$ is refined as $\eta$. We can note that $l_n^\text{min}=l_n^\text{max}$ if the bisection method converges. Thus, the optimal DNN partition point is derived as $l_n^*(t)=l_n^\text{min}$.

Given the optimized $l_n(t)$ and $P_m(t)$, we can rewrite (\ref{P3_1}) as\begin{alignat}{1}\label{P3_1_2}
	\min_{f_{m,n}^\text{G}(t), \forall n\in\mathcal{N}_m}\;&g_2(f_{m,n}^\text{G}(t))=\max_{n\in\mathcal{N}_m}\left\{K\tilde{D}_n\left(\frac{\sum_{l=1}^{l_n(t)}(o_l+o'_l)}{\phi_n^\text{D}f_n^\text{D}}\right.\right.\nonumber\\&\quad\quad\quad\quad\quad\!\left.\left.+\frac{\sum_{l=l_n(t)+1}^{L}(o_l+o'_l)}{\phi_m^\text{G}f_{m,n}^\text{G}(t)}\right)\right\}\\
	\text{s.t.}\quad &\textbf{C6}, \textbf{C10'}.\nonumber
\end{alignat}Similarly, the sub-problem in (\ref{P3_1_2}) can be solved by the bisection method. Let $g_2^\text{min}\!=\!K\min_{n\in\mathcal{N}_m}\!\{\tilde{D}_n\}\!\Big(\!\frac{\sum_{l=1}^{l_n(t)}(o_l+o'_l)}{\max_{n\in\mathcal{N}_m}\{\phi_n^\text{D}f_n^\text{D}\}}\!+\!\frac{\sum_{l=l_n(t)+1}^{L}(o_l+o'_l)}{\phi_m^\text{G}f_m,n^\text{G,max}}\!\Big)$ and $g_2^\text{max}\!=\!K\!\max\limits_{n\in\mathcal{N}_m}\!\{\tilde{D}_n\}\!\Big(\!\frac{\sum_{l=1}^{l_n(t)}(o_l+o'_l)}{\max_{n\in\mathcal{N}_m}\{\phi_n^\text{D}f_n^\text{D}\}}\!+\!\frac{\sum_{l=l_n(t)+1}^{L}(o_l+o'_l)}{\phi_m^\text{G}f_m,n^\text{G,min}}\!\Big)$ denote the lower and upper bound of $g_2(f_{m,n}^\text{G}(t))$. Let $\vartheta$ be the mid point of the interval $(g_2^\text{min}, g_2^\text{max})$, i.e., $\vartheta=\frac{1}{2}(g_2^\text{min}+g_2^\text{max})$. In each iteration, we first compute the lower bound of $f_{m,n}^\text{G}(t)$ according to $K\tilde{D}_n\Big(\frac{\sum_{l=1}^{l_n(t)}(o_l+o'_l)}{\phi_n^\text{D}f_n^\text{D}}+\frac{\sum_{l=l_n(t)+1}^{L}(o_l+o'_l)}{\phi_m^\text{G}f_{m,n}^\text{G}(t)}\Big)\leq \vartheta$, i.e., $f_{m,n}^\text{G}(t)\!\ge\! \big({\sum_{l=l_n(t)+1}^{L}(o_l\!+\!o'_l)}\big)/\phi_m^\text{G}/\Big(\!\!-\!\frac{\sum_{l=1}^{l_n(t)}(o_l+o'_l)}{\phi_n^\text{D}f_n^\text{D}}$ $+\frac{\vartheta}{K\tilde{D}_n}\Big)$. If constraints \textbf{C6} and \textbf{C10'} hold when $f_{m,n}^\text{G}(t)\!=\! {\big(\sum_{l=l_n(t)+1}^{L}(o_l+o'_l)\big)}/\phi_m^\text{G}/\!\Big(\!\!-\!\frac{\sum_{l=1}^{l_n(t)}(o_l+o'_l)}{\phi_n^\text{D}f_n^\text{D}}$ $+\frac{\vartheta}{K\tilde{D}_n}\Big)$, we refine the upper bound of $g_2(f_{m,n}^\text{G}(t))$ as $\vartheta$. Otherwise, the lower bound of $g_2(f_{m,n}^\text{G}(t))$ is refined as $\vartheta$. Suppose that $\vartheta=\vartheta^*$ when the bisection method converges. Thus, the optimal computation frequency is derived as $f_{m,n}^{G^*}(t)= \big({\sum_{l=l_n(t)+1}^{L}(o_l+o'_l)}\big)/{\phi_m^\text{G}/\Big(\frac{\vartheta^*}{K\tilde{D}_n}-\frac{\sum_{l=1}^{l_n(t)}(o_l+o'_l)}{\phi_n^\text{D}f_n^\text{D}}\Big)}$.

Given the optimized $l_n(t)$ and $f_{m,n}^\text{G}(t)$, we rewrite (\ref{P3_1}) as\begin{alignat}{1}\label{P3_1_3}
	\min_{P_m(t), \forall n\in\mathcal{N}_m}\;&g_3(P_m(t))=\frac{\gamma}{B^\text{u}\log_2\!\left(1\!+\!\frac{P_m(t)h^u_{m,j}(t)}{B^\text{u}N_0+i^u_{m,j}(t)}\right)}\\
	\text{s.t.}\quad &\textbf{C4},\textbf{C10'}.\nonumber
\end{alignat}Note that the sub-problem in (\ref{P3_1_3}) is convex. The optimal transmit power is as follows.
\begin{equation}
	\begin{split}
		\!\!\!\!\!\!\!\!P_m^*(t)\!=\!
		\begin{cases}
			&\!\!\!\!0,\;\mbox{if}~\frac{B^\text{u}}{\gamma\ln2}\!\!\left(\!E_m^\text{G}(t)-\!\!\!\!\!\sum\limits_{n\in\mathcal{N}_m}\!\!K\tilde{D}_n\frac{v_m^\text{G}}{\phi_m^\text{G}}\!\left(\sum\limits_{l=l_n(t)+1}^L\right.\right.\!\!\!\!\!\!\!\!\\&\;\;(o_l\!+\!o'_l)\!\bigg)(f_{m,n}^\text{G}(t))^2\!\Bigg)\!-\!\frac{BN_0+i^u_{m,j}(t)}{h^u_{m,j}(t)}\leq 0,\!\!\!\!\!\!\!\!\\
			&\!\!\!\!\min\{x^*,P_m^\text{max}\},\;\mbox{otherwise},
		\end{cases}
	\end{split}
\end{equation}where $x^*>0$ is the solution to $\frac{B^\text{u}}{\gamma}\Big(E_m^\text{G}(t)\!-\!\!\sum_{n\in\mathcal{N}_m}\!\!K\tilde{D}_n$ $v_m^\text{G}$ $\frac{\sum_{l=l_n(t)+1}^L(o_l+o'_l)}{\phi_m^\text{G}}(f_{m,n}^\text{G}(t))^2\Big)\!\log_2\!\Big(1+\frac{h^u_{m,j}(t)x}{B^\text{u}N_0+i^u_{m,j}(t)}\Big)-x=0$.

\subsubsection{Optimal channel assignment}
Given the optimized $\boldsymbol{\Lambda}(t)$, the channel assignment matrix $\boldsymbol{I}(t)$ can be optimized as\begin{alignat}{1}\label{P3_2}
	\min_{\boldsymbol{I}(t)}\;&V\!\max_{m\in\mathcal{M}}\left\{\sum_{j\in\mathcal{J}}I_{m,j}(t)\Lambda_{m,j}(t)\!\right\}\!-\!\!\!\sum_{m\in \mathcal{M}}\sum_{j\in\mathcal{J}}Q_m(t)I_{m,j}(t)\\
	\text{s.t.}\quad &\textbf{C1}\sim\textbf{C3}.\nonumber
\end{alignat}To solve the problem in (\ref{P3_2}), we first introduce an auxiliary variable $\lambda$, and thus the problem can be equivalently transformed into\begin{alignat}{1}\label{P3_2_1}
&\min_{\lambda,\boldsymbol{I}(t)}\;\lambda-\sum_{m\in \mathcal{M}}\sum_{j\in\mathcal{J}}Q_m(t)I_{m,j}(t)\\
\text{s.t.}\quad
&\textbf{C1}\sim\textbf{C3},\;\textbf{C12}:\lambda\ge V\sum_{j\in\mathcal{J}}I_{m,j}(t)\Lambda_{m,j}(t), \forall m\in\mathcal{M}.\nonumber
\end{alignat}Following the solution of the problem in (\ref{P3_1}), we decompose (\ref{P3_2_1}) into two sub-problems in (\ref{R4}) and (\ref{R8}), and then optimize the auxiliary variable $\lambda$ and the channel assignment matrix $\boldsymbol{I}(t)$ by solving the sub-problems in an iterative manner.

Given the optimized auxiliary variable $\lambda$, we can rewrite (\ref{P3_2_1}) as\begin{alignat}{1}\label{R4}
	&\min_{\boldsymbol{I}(t)}\quad-\sum_{m\in \mathcal{M}}\sum_{j\in\mathcal{J}}Q_m(t)I_{m,j}(t)\\
	\text{s.t.}\quad&\textbf{C1}:I_{m,j}(t)\in \{0,1\},\forall m\in \mathcal{M},j\in \mathcal{J},\nonumber\\&\textbf{C2}:\sum_{j\in\mathcal{J}}I_{m,j}(t)\leq 1, \forall m\in \mathcal{M},\nonumber\\&\textbf{C3}:\sum_{m\in\mathcal{M}}I_{m,j}(t)=1, \forall j\in \mathcal{J},\nonumber\\&\textbf{C12}:\sum_{j\in\mathcal{J}}V\Lambda_{m,j}(t)I_{m,j}(t)\leq\lambda, \forall m\in\mathcal{M}.\nonumber
\end{alignat}From constraints \textbf{C1} and \textbf{C2}, constraint \textbf{C12} can be equivalently transformed into \textbf{C12'}, i.e., $I_{m,j}(t)=0, \forall (m, j)\in \left\{(m, j)\in \mathcal{M}\times\mathcal{J}\vert \Lambda_{m,j}(t)>\lambda/V\right\}$. By replacing the corresponding weights $Q_m(t)$ in the objective function of (\ref{R4}) with an extremely large positive value, the problem in (\ref{R4}) can be transformed into a standard weighted bipartite matching linear program as\begin{alignat}{1}\label{R6}
	&\min_{\boldsymbol{I}(t)}\quad\sum_{m\in \mathcal{M}}\sum_{j\in\mathcal{J}}\Theta_{m,j}I_{m,j}(t)\\
	\text{s.t.}\quad&\textbf{C1}:I_{m,j}(t)\in \{0,1\},\forall m\in \mathcal{M},j\in \mathcal{J},\nonumber\\&\textbf{C2}:\sum_{j\in\mathcal{J}}I_{m,j}(t)\leq 1, \forall m\in \mathcal{M},\nonumber\\&\textbf{C3}:\sum_{m\in\mathcal{M}}I_{m,j}(t)=1, \forall j\in \mathcal{J},\nonumber
\end{alignat}where\begin{equation}
	\begin{split}
		\!\!\!\!\!\Theta_{m,j}\!=\!
		\begin{cases}
			\Psi,\,\mbox{if}~\!(m, j)\!\in\!\left\{(m, j)\!\in\! \mathcal{M}\!\times\!\mathcal{J}\vert V\!\Lambda_{m,j}(t)\!>\!\lambda\right\},\!\!\!\!\!\!\!\!\!\!\!\\
			-Q_m(t),\,\mbox{otherwise}.
		\end{cases}
	\end{split}
\end{equation}Note that $\Psi$ is set as an extremely large positive value to create the composite objective function in (\ref{R6}) which incorporates the effect of constraint \textbf{C12'}. Based on the Hungarian method\cite{DBLP:books/daglib/0022248,0On}, the optimal channel assignment matrix $\boldsymbol{I}^*(t)$ can be obtained in polynomial time.

Given the optimized channel assignment matrix $\boldsymbol{I}(t)$, we can rewrite (\ref{P3_2_1}) as\begin{alignat}{1}\label{R8}
	&\quad\quad\quad\quad\quad\quad\quad\quad\min_{\lambda}\quad\lambda\\
	&\text{s.t.}\quad \textbf{C12}:\lambda\ge V\sum_{j\in\mathcal{J}}I_{m,j}(t)\Lambda_{m,j}(t), \forall m\in\mathcal{M}.\nonumber
\end{alignat}Obviously, the optimal auxiliary variable is given by\begin{alignat}{1}\lambda^*=\max_{m\in\mathcal{M}}\left\{\sum_{j\in\mathcal{J}}I_{m,j}(t)\Lambda_{m,j}(t)\right\}.
\end{alignat}

With the optimal channel assignment matrix $\boldsymbol{I}^*(t)$, whether the $m$-th gateway and its associated devices are selected to participate in the local model training in the $t$-th communication round is determined by $\mathbbm{1}_m^t=\sum_{j\in\mathcal{J}}i_{m,j}(t)$.

\subsection{Optimality, Complexity, Applicability, and Scalability Analysis}
In this subsection, we present the optimality, complexity, applicability, and scalability analysis of the proposed DDSRA algorithm as follows.

\textit{Optimality analysis:} The DDSRA algorithm converges to at least a locally optimal solution. From Section V-B, the DDSRA algorithm consists of two parts: a) solve the auxiliary variables $\Lambda_{m,j}(t)$ in (\ref{P3_1}) based on block coordinate descent method in the outer layer loop and bisection method in the inner layer loop, and b) solve the channel assignment matrix $\boldsymbol{I}(t)$ in (\ref{P3_2_1}) based on block coordinate descent method in the outer layer loop and Hungarian method in the inner layer loop. According to the existing works on the block coordinate descent method\cite{DBLP:journals/tvt/FuMWY21,DBLP:journals/tvt/HuaWWDHY19,DBLP:journals/twc/WuZZ18}, the convergence to a local optimum can be guaranteed when the sub-problems in each iteration can be solved exactly with optimality. Notably, bisection method is an efficient and widely-used algorithm which can converge to the global optimum superlinearly\cite{Phillips2001,8314727}, and Hungarian method is a straightforward method of finding the optimal solution to an assignment problem \cite{DBLP:books/daglib/0022248,0On}. According to the above analysis, the DDSRA algorithm converges to at least a locally optimal solution.
	
\textit{Complexity analysis:} Let $L_1$ and $L_2$ denote the required number of iterations for solving (\ref{P3_1}) based on the block coordinate descent method in the outer layer loop and the bisection method in the inner layer loop, respectively. The computational complexity of solving the auxiliary variables $\Lambda_{m,j}(t)$ is represented as $\mathcal{O}(NJL_1L_2)$. From Section V-B, the computational complexity of the Hungarian method in the inner layer is represented as $\mathcal{O}(M^3)$. Let $L_3$ denote the required number of iterations for solving (\ref{P3_2_1}) based on the block coordinate descent method in the outer layer loop. The computational complexity of solving the channel assignment matrix $\boldsymbol{I}(t)$ in (\ref{P3_2_1}) is represented as $\mathcal{O}(M^3L_3)$. To sum up, the total computational complexity of DDSRA is $\mathcal{O}(NJL_1L_2+M^3L_3)$.
	
\textit{Applicability analysis:} The DDSRA algorithm is applicable to a variety of IIoT scenarios. Thanks to the joint optimization of device scheduling and resource allocation (i.e. DNN partition point, channel assignment, transmit power, and computation frequency), our proposal can be potentially applied in device heterogeneity scenarios wherein IIoT devices are intrinsically heterogeneous in computational capacity and memory resource. Thanks to the developed device-specific participation rate linked to the training dataset size and data distribution, our proposal is robust against data heterogeneity (i.e., non-IID data distribution). Moreover, thanks to the layer-level memory usage and FLOPs calculation model, our proposal is applicable to diverse DNN models such as multilayer perceptron (MLP) and convolutional neural network (CNN).
	
\textit{Scalability analysis:} The computational complexity of the DDSRA algorithm is directly proportional to the number of end devices, i.e., $N$. Furthermore, by exploiting the independence between the auxiliary variables $\Lambda_{m,j}(t)$, the DDSRA algorithm is decomposed into $MJ$ multi-threaded parallel computation tasks (see line 5 in \textbf{Algorithm} \ref{alg::Dynamic Resource Allocation and Client Scheduling Algorithm}), which can greatly reduce the time complexity. Therefore, the DDSRA algorithm is scalable to a large number of end devices.

\section{Performance Analysis}\label{sec:Performance Analysis}
\subsection{Asymptotic Optimality of DDSRA}
In this subsection, we will analyze the performance of the proposed DDSRA algorithm in terms of asymptotic optimality, and characterize the trade-off between the delay minimization and the degree of which the participation rate constraint is satisfied.
\begin{theorem}\label{performance analysis}
	With the optimal policy of \rm{\textbf{P2}} in each communication round, and note that $\mathbb{E}\{Q(0)\}<\infty$, we have
	\begin{alignat}{1}\label{equ2:theorem1}
		\varphi^*-\varphi^\text{opt}\leq \frac{H}{V}+\frac{\mathbb{E}\{\Xi(0)-\Xi(T)\}}{VT},\end{alignat}and
	\begin{equation}\label{equ3:theorem1}
			\frac{1}{T}\!\sum_{t=0}^{T-1}\!\mathbbm{1}_m^t\ge \Gamma_m-\!\sqrt{ \frac{H\!+\!V(\varphi^\text{opt}\!-\!\tau^\text{min})}{T}+\!\!\sum_{m\in \mathcal{M}}\!\!\frac{\mathbb{E}\{Q_m(0)^2\}}{T^2}},
	\end{equation}where $\varphi^\text{opt}$ is the optimal utility of \textbf{P0} over all possible scheduling policies, $\varphi^*$ represents the optimal utility of \textbf{P2}, and $\tau^\text{min}\!=\!\frac{K\min_{n\in\mathcal{N}}\left\{\tilde{D}_n\right\}\sum_{l=1}^L(o_l+o'_l)}{\min\{\min_{n\in\mathcal{N}}\{\phi_n^\text{D}f_n^\text{D}\},\min_{m\in\mathcal{M}}\{\phi_m^\text{G}f_m^\text{G,max}\}\}}\!+\!{\gamma}/B^\text{u}/\log_2\big(1$ $+ {P_m^\text{max}\overline{h^u_{m,j}}}/\big({B^\text{u}N_0\!+\!\overline{i^u_{m,j}}}\big)\big)\!+\!{\gamma}/B^\text{d}/\!\log_2\!\big(1\!+\!{P^B\overline{h^d_{m,j}}}/\big(B^\text{d}$ $N_0\!+\!\overline{i^d_{m,j}}\big)\big)$.
\end{theorem}
\begin{IEEEproof}
	Please see Appendix C.
\end{IEEEproof}

We have verified the asymptotic optimality of the proposed DDSRA algorithm in (\ref{equ2:theorem1}). That is, the proposed DDSRA algorithm converges to the optimal solution as $V$ increases. Moreover, (\ref{equ3:theorem1}) indicates that the participation rate of each gateway and its associated devices increases, and finally converges to the optimized device-specific participation rate $\Gamma_m$ as $V$ decreases. Hence, \textbf{Theorem} \ref{performance analysis} shows an [$\mathcal{O}(1/V)$, $\mathcal{O}(\sqrt{V})$] trade-off between the minimization of FL training latency and the degree of which the participation rate constraint is satisfied, where the control parameter $V$ represents how much we emphasize the maximization of the FL training latency. To be specific, a large value of $V$ encourages reducing the FL training latency, which can be adopted for real-time delay-sensitive IIoT applications. Meanwhile, a small value of $V$ pushes the participation rate of each gateway and its associated devices to the optimized device-specific participation rate $\Gamma_m$, thereby promoting the FL training performance.


\subsection{FL Convergence Analysis of the DDSRA Algorithm}
For ease of exposition, we define $\delta= \max_n \{\delta_n\}$, $\sigma= \max_n \{\sigma_n\}$, $F_n(\boldsymbol{w})=f(\boldsymbol{w},\mathcal{D}_n)$, $\tilde{F}_n(\boldsymbol{w})=f(\boldsymbol{w},\tilde{\mathcal{D}}_n)$, and $F(\boldsymbol{w})=f(\boldsymbol{w},\cup\mathcal{D}_n)$. Based on \textbf{Assumption} \ref{assumption3} and \ref{assumption4}, the FL convergence bound of the proposed algorithm is derived as follows.
\begin{theorem}\label{theorem2}
	Assume that the loss function $F_n(w)$ is convex, $L_n$-smooth and $\rho_n -$Lipschitz continuous, the FL convergence bound is represented as
	\begin{alignat}{1}\label{theorem2_1}
		&\mathbb{E}\left[F\left(\boldsymbol{W}^{T}\right)-F\left(\boldsymbol{w}^*\right)\right]\leq\nonumber\\&\frac{1}{T\!\left(\!\beta\phi\!-\!\frac{\rho\left(\!\delta+\!\!\!\sum\limits_{n\in\mathcal{N}}\!\!\xi_n\!\!\frac{\sigma_n}{\sqrt{\tilde{D}_n}}\!\right)\!\left((\beta L+1)^K\!-\!1\right)+\beta\left(\!\delta+\!\!\!\sum\limits_{n\in\mathcal{N}}\left\vert\xi_n\!-\!\frac{D_n}{\sum\limits_{n\in\mathcal{N}}\!\!D_n}\right\vert\rho_n\!\right)}{\varepsilon^2KL}\!\right)},
	\end{alignat}where $L=\max_n \{L_n\}$, $\rho=\max_n\{\rho_n\}$, $\phi\triangleq\omega(1$ $-{\beta L}/{2})$, $\omega\triangleq \min_{t\in\mathcal{T}} \frac{1}{\left\Vert \boldsymbol{v}^{K,t-1}-\boldsymbol{w}^*\right\Vert^2}$, $\xi_n=\frac{\sum_{m\in\mathcal{M}} \Gamma_ma_{m,n}\tilde{D}_n}{\sum_{n\in\mathcal{N}}\sum_{m\in\mathcal{M}} \Gamma_ma_{m,n}\tilde{D}_n}$, and $\varepsilon\triangleq \min_{t\in\mathcal{T}}\left[ F(\boldsymbol{v}^{K,t})-F(\boldsymbol{w}^*)\right]$.
\end{theorem}
\begin{IEEEproof}
	Please see Appendix D.
\end{IEEEproof}From \textbf{Theorem} \ref{theorem2}, we can see that larger training data sizes $\tilde{D}_n$ can reduce the value of the term $\rho\Big(\delta+\sum\limits_n\xi_n\frac{\sigma_n}{\sqrt{\tilde{D}_n}}\Big)\left((\beta L+1)^K-1\right)$ in (\ref{theorem2_1}), which contributes to a smaller convergence bound and thereby a better FL performance. In addition, by setting a larger participation rate for important devices with better data distribution (i.e., lower variances $\sigma_n$), our derived device-specific participation rate $\Gamma_m$ can produce a lower weighted sum of $\frac{\sigma_n}{\sqrt{\tilde{D}_n}}$, thereby leading to a lower FL convergence bound. Moreover, it also shows that when the participation rate is set to be the same for each gateway and its associated devices, i.e., $\Gamma_m = \Gamma_{m'}, \forall m\neq m'$, the term $\left\vert\xi_n-\frac{D_n}{\sum_nD_n}\right\vert$ in (\ref{theorem2_1}) is zero if the training data sizes are proportionate to the local dataset sizes (i.e., $\tilde{D}_n=\alpha D_n$, $\forall n\in\mathcal{N}$, where $\alpha$ represents the data sampling ratio for the local datasets). That is, the training data sizes proportionate to the local dataset sizes can lead to a small convergence bound, thereby promoting a better FL performance.


In addition, we derive a convergence bound for non-convex setting in \textbf{Theorem} \ref{theorem3}. \textbf{Theorem} \ref{theorem3} below indicates that the proposed DDSRA algorithm can achieve a FL convergence rate of $\mathcal{O}(1/T)$ for non-convex loss functions.
\begin{theorem}\label{theorem3}
	Assume that the loss function $F_n(w)$ is non-convex and $L_n$-smooth, the convergence bound is represented as\begin{alignat}{1}\label{theorem3_1}
		&\frac{1}{T}\!\sum_{t=0}^{T-1}\!\mathbb{E}\!\left[\left\Vert\nabla F\!\left(\tilde{\boldsymbol{w}}^{t}\right)\right\Vert^2\right]\!\leq\! \frac{2}{K\beta T}\left(\mathbb{E}\!\left[F\!\left(\tilde{\boldsymbol{w}}^{0}\right)\right]\!-\!\mathbb{E}\!\left[F\!\left(\tilde{\boldsymbol{w}}^T\right)\right]\right)+\nonumber\\&\frac{L\beta N}{T}\!\sum_{t=0}^{T-1}\sum_{n=1}^{N}\sum_{k=0}^{K-1}\!\!\left(\!\frac{\sum_{m\in\mathcal{M}} \Gamma_ma_{m,n}\tilde{D}_n}{\sum_{n\in\mathcal{N}}\sum_{m\in\mathcal{M}} \Gamma_ma_{m,n}\tilde{D}_n}\!\right)^{\!2}\!\mathbb{E}\Big[\Vert\nabla F_n\big(\nonumber\\&\tilde{\boldsymbol{w}}_n^{k,t}\big)\big\Vert^2\Big]\!+\!\frac{N\beta^2}{KT}\sum_{t=0}^{T-1}\sum_{n=1}^{N}\!\sum_{k=0}^{K-1}\!\!\left(\!\frac{\sum_{m\in\mathcal{M}} \Gamma_ma_{m,n}\tilde{D}_n}{\sum_{n\in\mathcal{N}}\sum_{m\in\mathcal{M}} \Gamma_ma_{m,n}\tilde{D}_n}\!\right)^{\!2}\nonumber\\&L_n^2\beta^2k\sum_{j=0}^{k-1}\mathbb{E}\left[\left\Vert\nabla F_n\left(\tilde{\boldsymbol{w}}_n^{j,t}\right)\right\Vert^2\right].
	\end{alignat}
\end{theorem}
\begin{IEEEproof}
	Please see Appendix E.
\end{IEEEproof}

\section{Experiential Results}\label{sec:Experiential Results}
\subsection{Experimental Setting}
To evaluate the FL training performance of the proposed algorithm for complex datasets and DNNs, we utilize Street View House Numbers (SVHN)\cite{DBLP:conf/icpr/SermanetCL12} and CIFAR-10\cite{DBLP:conf/iscas/YangBMVM18} datasets trained on VGG-11\cite{DBLP:journals/eswa/SharmaS21a} for non-IID setting to demonstrate the test accuracy performance.
\begin{itemize}
	\item \textbf{SVHN}. SVHN contains over $600000$ $32\times32$ RGB images in $10$ classes (from 0 to 9), which is cropped from pictures of house number plates.
	\item \textbf{CIFAR}-10. The CIFAR-10 dataset consists of $60000$ $32\times32$ RGB images in $10$ classes (from 0 to 9), with $50000$ training images and $10000$ test images per class.
\end{itemize}
For non-IID setting, we follow the previous work\cite{DBLP:journals/corr/abs-1806-00582} to distribute the data points in each local dataset. The data points are sorted by class and divided into two extreme cases: (a) $q_m$-class non-IID, where each device holds data points in $q_m$ classes, and (b) IID, where each device holds data points in all of the $10$ classes. In this experiment, $q_m$ is randomly generated, and we set the non-IID degree of the data distribution (proportion of the $q_m$-class non-IID data points) as $\chi=1$.

For comparison purpose, we also consider the following baseline schemes:
\begin{itemize}
	\item \textbf{Random Scheduling}\cite{DBLP:journals/tcom/YangLQP20}. The BS uniformly selects $J$ gateways and the associated devices at random for local model training in each communication round.
	\item \textbf{Round Robin}\cite{DBLP:journals/tcom/YangLQP20}. The BS divides the $M$ gateways and the associated devices into $\lceil \frac{J}{M}\rceil$ groups and consecutively assigns each group to the wireless channels in each communication round.
	\item \textbf{Loss Driven Scheduling}. The BS selects $J$ gateways and the associated devices according to the local training loss for local model update in each communication round.
	\item \textbf{Delay Driven Scheduling}. The BS selects $J$ gateways and the associated devices for local model training with the objective of minimizing FL latency in each communication round.
\end{itemize}

Besides, we consider $M=6$ gateways, $N=12$ devices, and $J=3$ channels. Each gateway is designed to be associated with $2$ of the devices. For each device, the local dataset size $D_n$ is uniformly distributed within $(0,2000]$, $E_n^\text{D,max}=5$ J, $G_n^\text{D,max}=2$ GB, $f_n^\text{D}$ is uniformly distributed within $[0.1,1]$ GHz, $\phi_n^\text{D}=16$ FLOPs per CPU cycle\cite{DBLP:journals/tjs/Dolbeau18}, and $v_n^\text{D}=10^{\text{-27}}$. For each gateway, $d_m$ is uniformly distributed within $[1000,2000]$ m, $E_m^\text{G,max}=30$ J, $G_m^\text{G,max}=4$ GB, $f_m^\text{G,max}=4$ GHz, $\phi_m^\text{D}=32$ FLOPs per CPU cycle, $v_m^\text{G}=10^{\text{-27}}$, and $P_m^\text{max}=200$ mW. The channel parameters are set as $d_0=1$ m, $\nu=2$, $B^\text{u}=1$ MHz, $B^\text{d}=20$ MHz, $N_0=-174$ dBm/Hz, $h_0=-30$ dB, $P^\text{BS}=1$ W, the uplink and downlink interferences $i_{m,j}^\text{u}(t)$ and $i_{m,j}^\text{d}(t)$ are produced by the Gaussian distribution with different variances, and the channel power gains $\rho_{m,n}^\text{u}(t)$ and $\rho_{m,n}^\text{d}(t)$ are exponentially distributed with unit mean. For local model training, we set local epoch $K=5$, training data sampling ratio $\alpha=0.05$, and learning rate $\beta=0.01$. The memory usage and FLOPs for the DNN layers trained at the device and gateway side can be calculated according to Table \ref{table1}. In addition, the values of $L_n$, $\sigma_n$, $\delta_n$ and $\rho_n$ are estimated by observing the model parameters in the FL training process.

\subsection{Performance of Device-specific Participate Rate Policy}
\begin{figure}[!t]
	\centering
	\subfigure[]{\includegraphics[width=3.2in]{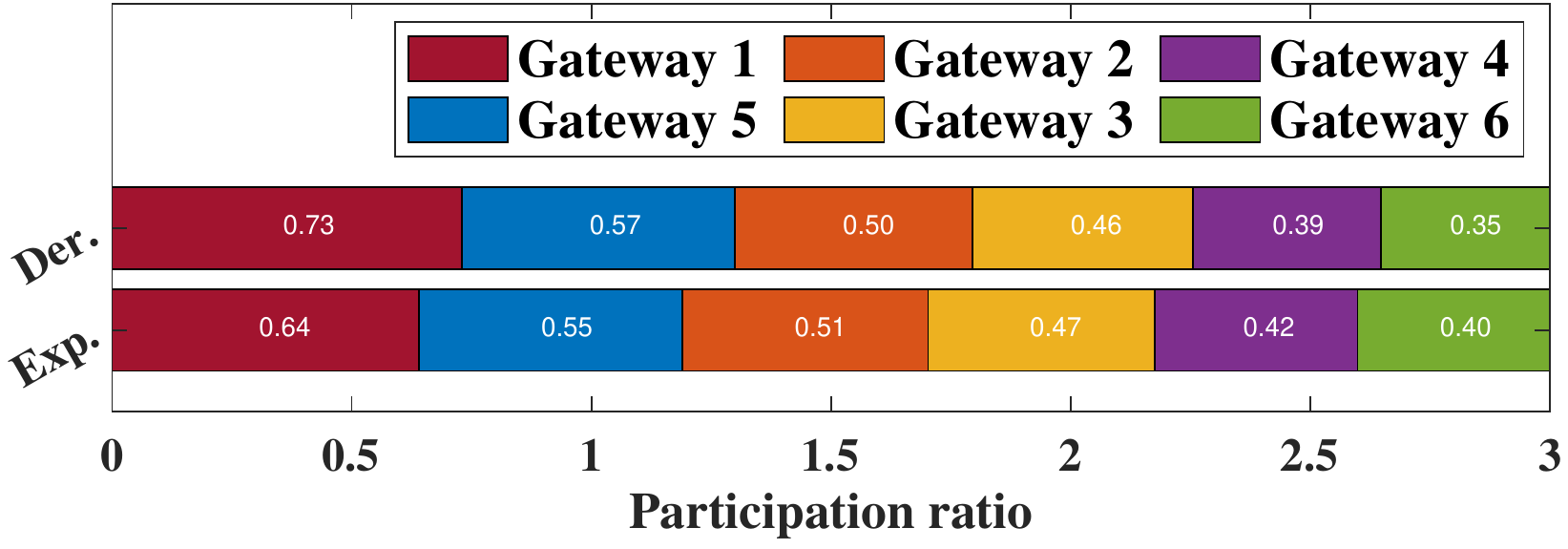}}
	\subfigure[]{\includegraphics[width=3.2in]{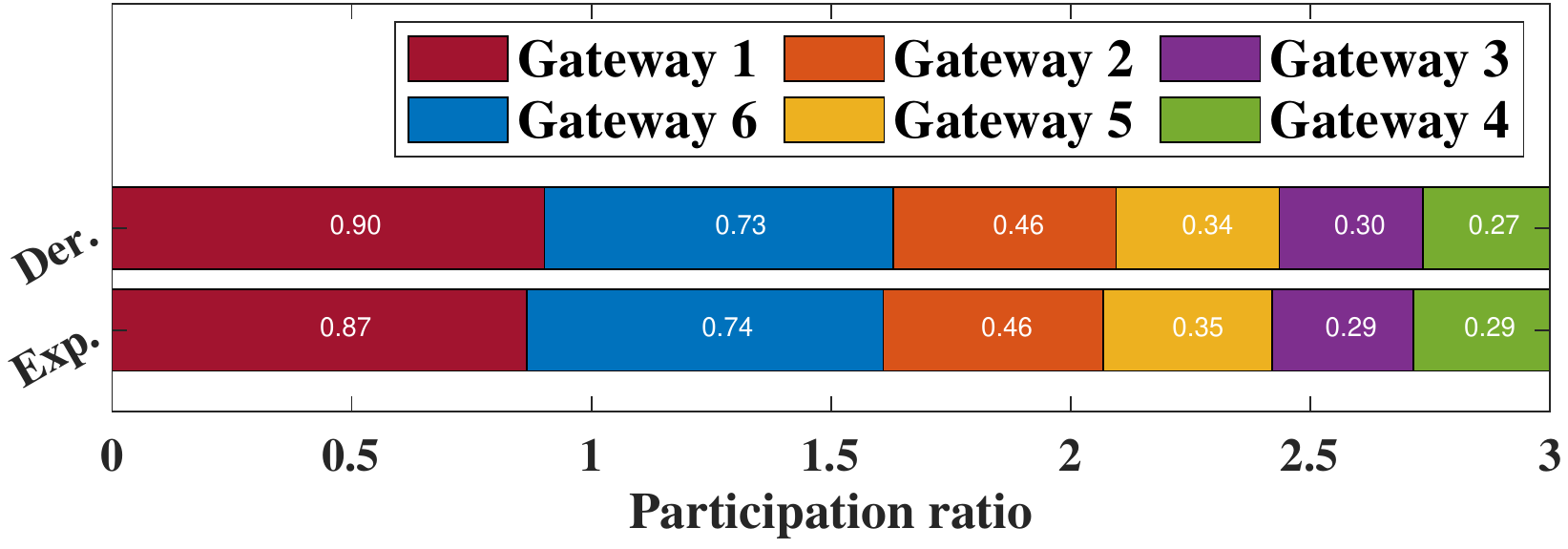}}
	\caption{The derived and experimental participation rate of each gateway and associated devices on (a) SVHN and (b) CIFAR-10 datasets.}
	\label{fig:11}
\end{figure}
\begin{figure}[!t]
	\centering
	\subfigure[]{\includegraphics[width=3.2in]{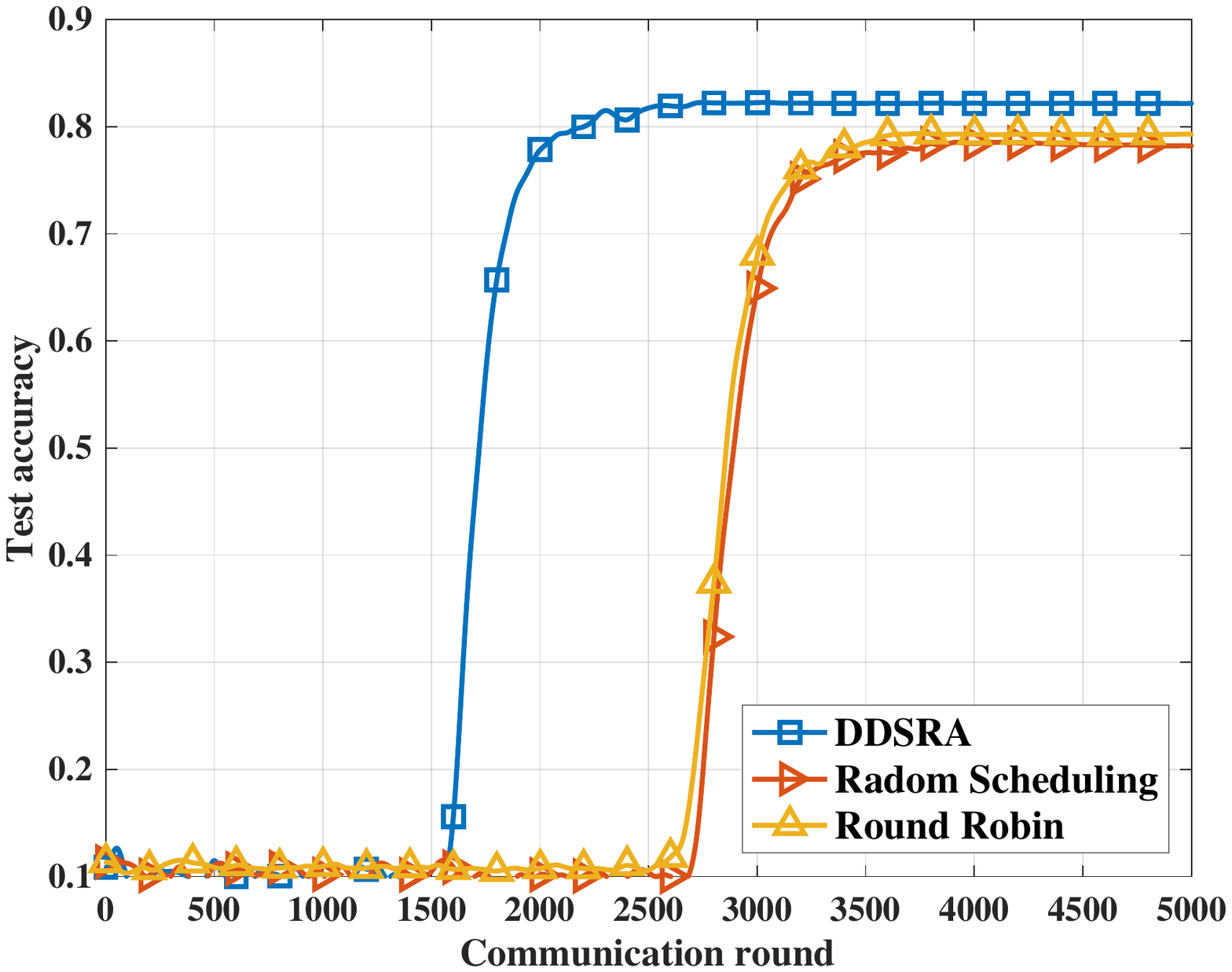}}
	\subfigure[]{\includegraphics[width=3.2in]{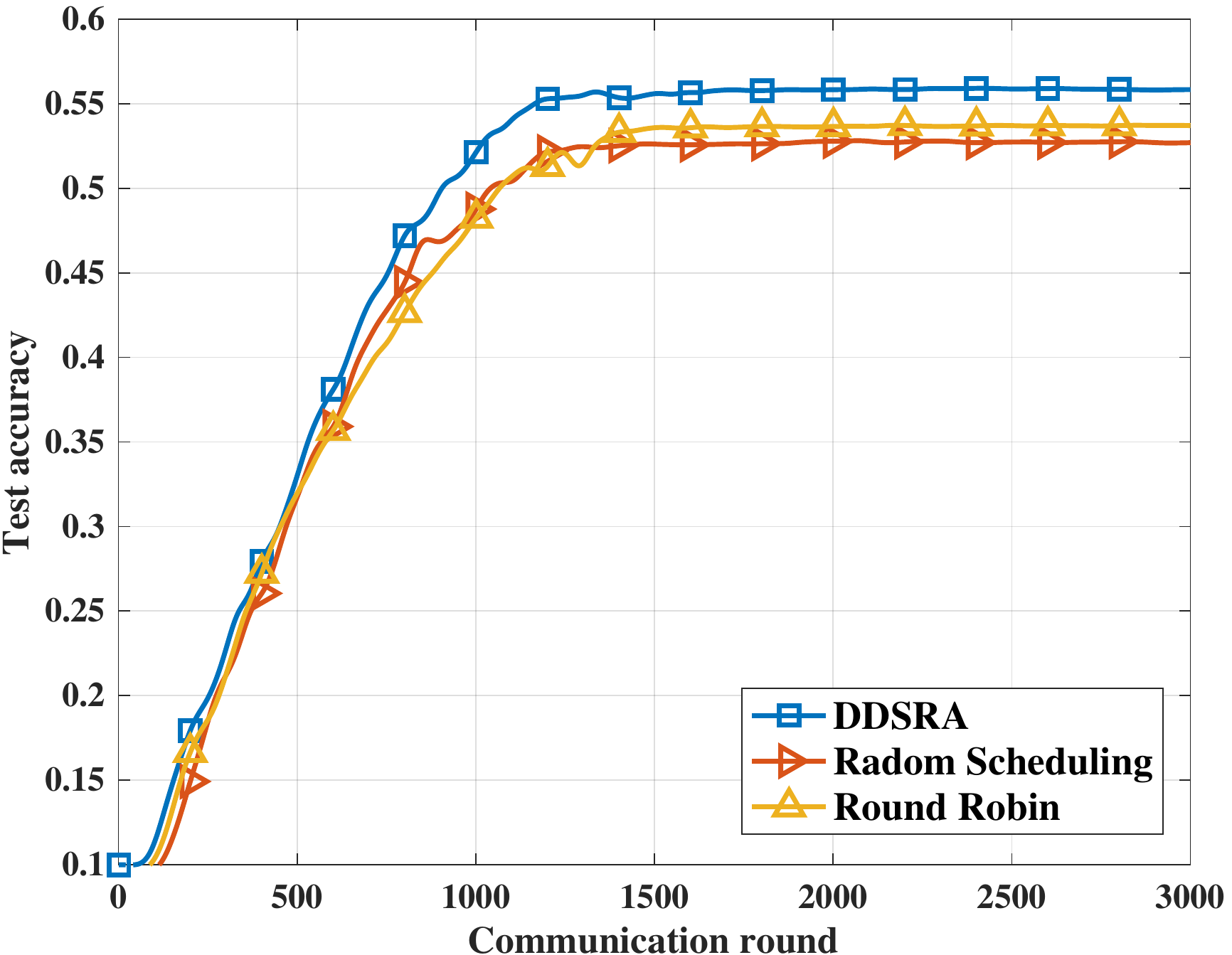}}
	\caption{Test accuracy comparison between the proposed device-specific participation rate policy and the baseline device scheduling policies (i.e., Random Scheduling policy and Round Robin policy) on (a) SVHN and (b) CIFAR-10 datasets.}
	\label{fig:12}
\end{figure}
To demonstrate the derived device-specific participation rate linked to FL performance in Section \ref{sec:Device-specific measurement of the participation rate}, we compare our derived participation rate of each gateway and the associated devices in (\ref{12}) with the experimental value on SVHN and CIFAR-10 datasets, as shown in Fig.\ref{fig:11}. Note that the derived value is calculated based on the upper bound of the divergence between the local model parameters learned in the FL training process and the model parameters learned in the centralized training process, i.e., $\vert\vert\hat{\boldsymbol{w}}_m^t-\boldsymbol{v}^{K,t}\vert\vert$, while the experimental value is obtained by observing $\vert\vert\hat{\boldsymbol{w}}_m^t-\boldsymbol{v}^{K,t}\vert\vert$ in the training process. First, Fig.\ref{fig:11} shows that the derived participation rate of each gateway and the associated devices is consistent with the experimental value, which justifies the divergence bound in \textbf{Theorem} \ref{theorem1}. Second, Fig.\ref{fig:11} shows that the $1$-th gateway and the associated devices can achieve the highest  participation rate. This is because we set each device associated with the $1$-th gateway a local dataset with a wider variety of the $q_m$-class non-IID data points, which makes the data distribution of the devices associated with the $1$-th gateway better represents the overall data distribution. In addition, Fig.\ref{fig:12} shows the comparison of the test accuracy between the proposed device-specific participation rate policy, Random Scheduling policy and Round Robin policy on SVHN and CIFAR-10 datasets. It can be observed that, with the same number of participant gateways and associated devices in each communication round, the proposed device-specific participation rate policy achieves better learning performance than the baseline schemes with fairness guarantee. Compared with Random Scheduling policy, the proposed device-specific participation rate policy reduces the number of communication rounds required for convergence by $35\%$ for SVHN dataset, and improves the test accuracy by $6\%$ for CIFAR-10 dataset.

\subsection{Performance of DDSRA Algorithm}
\begin{figure}[!t]
	\centering
	\subfigure[]{\includegraphics[width=3.2in]{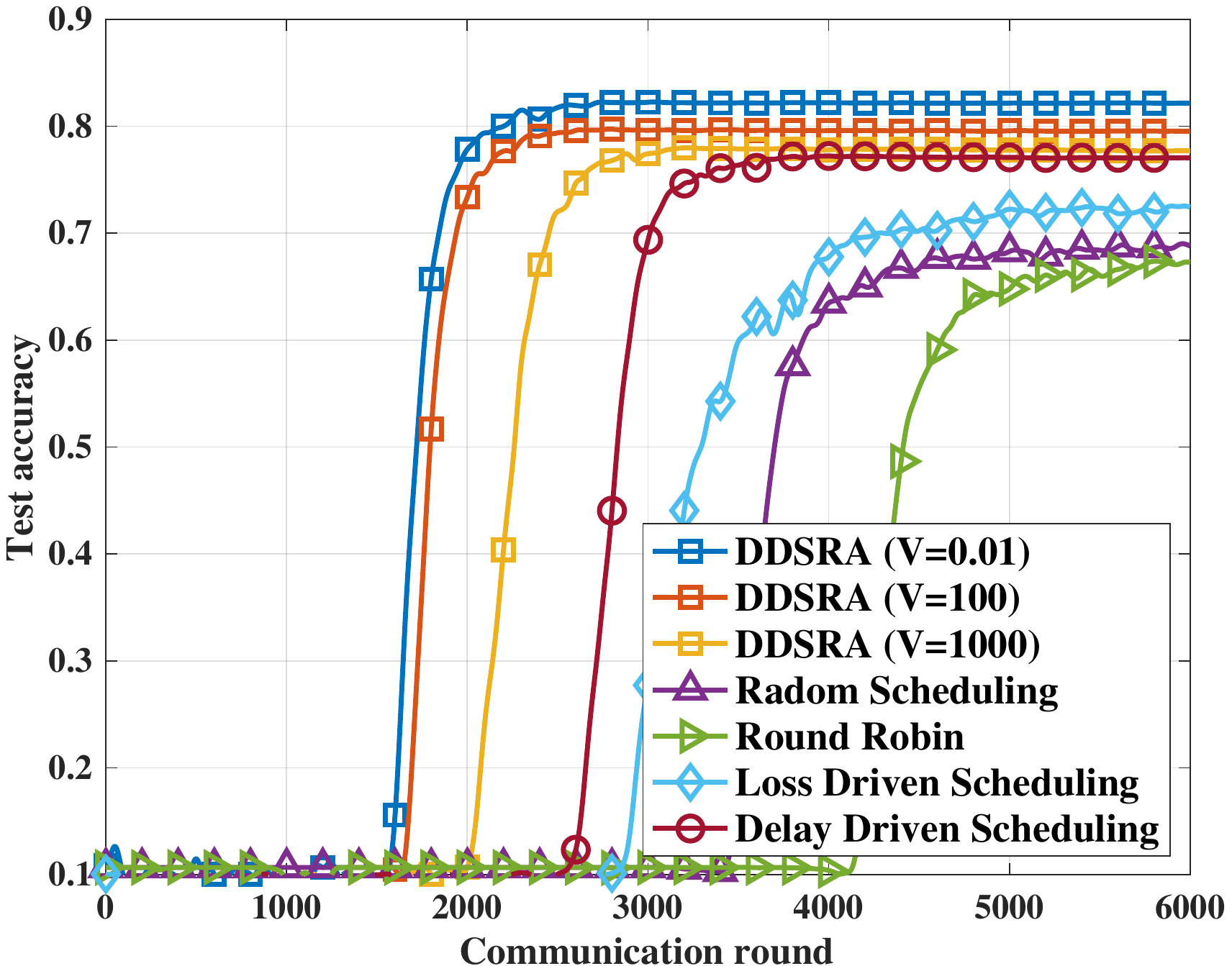}}
	\subfigure[]{\includegraphics[width=3.2in]{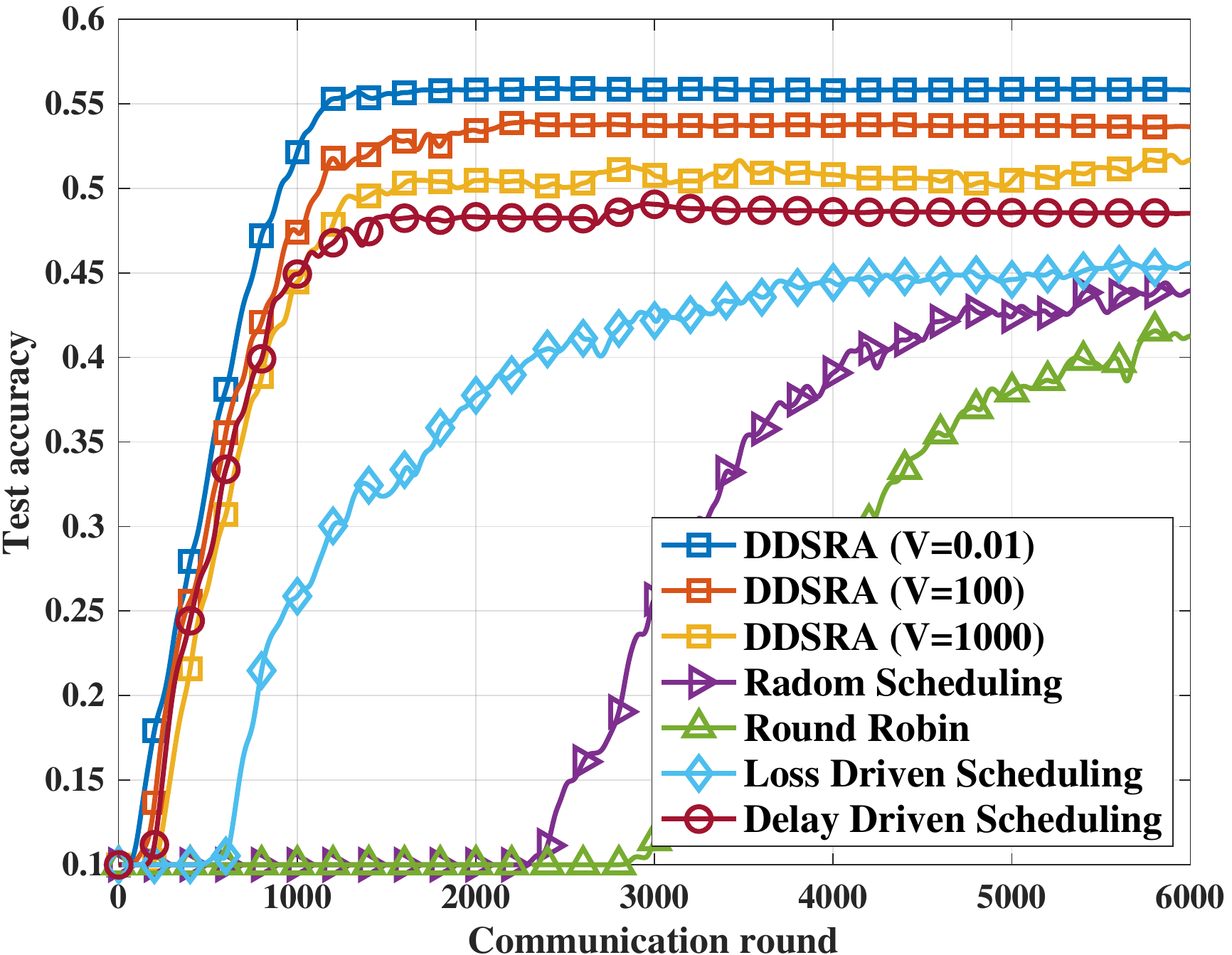}}
	\caption{Test accuracy comparison between DDSRA algorithm and the baseline schemes on (a) SVHN and (b) CIFAR-10 datasets.}
	\label{fig:13}
\end{figure}
\begin{figure}[!t]
	\centering
	\subfigure[]{\includegraphics[width=3.2in]{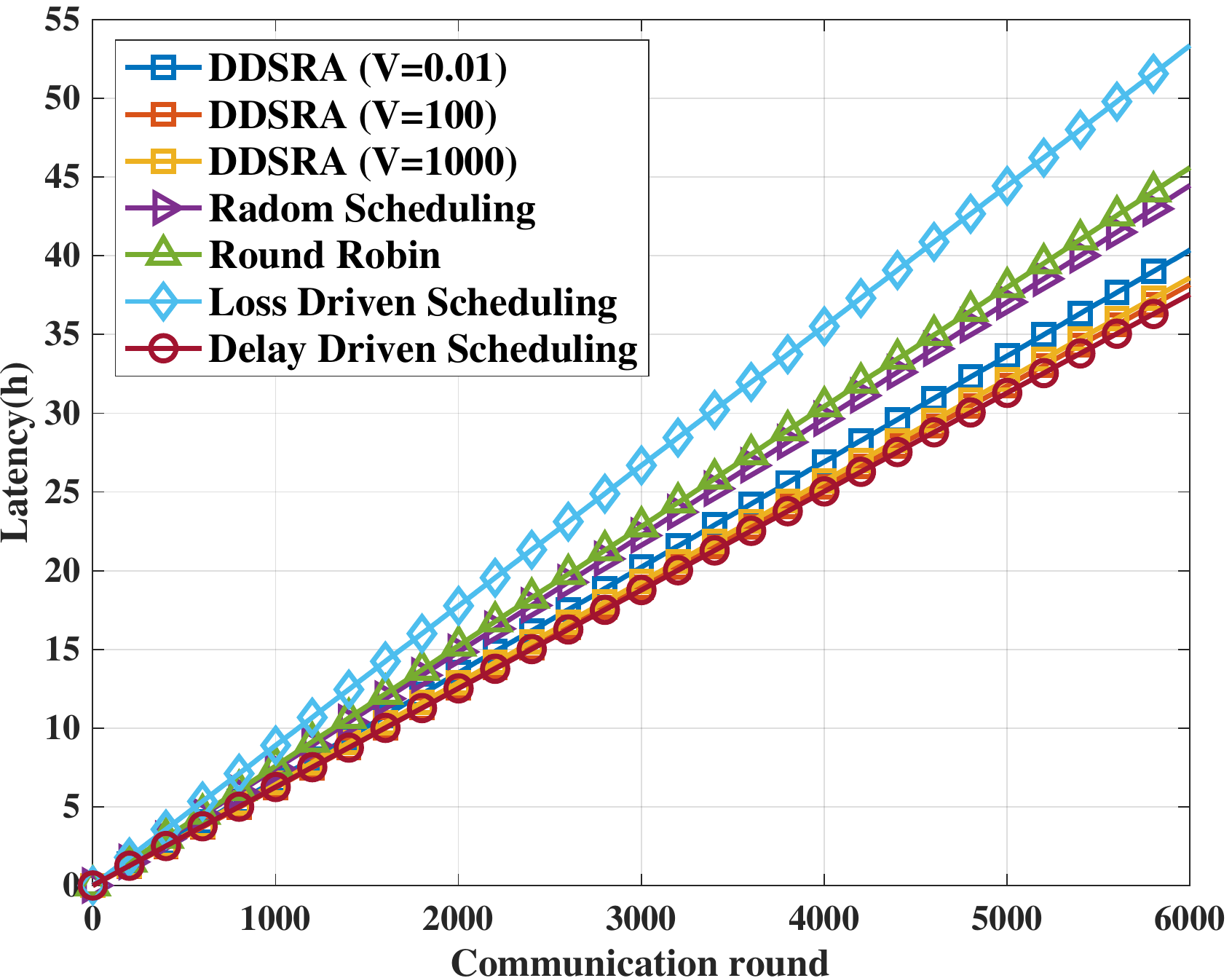}}
	\subfigure[]{\includegraphics[width=3.2in]{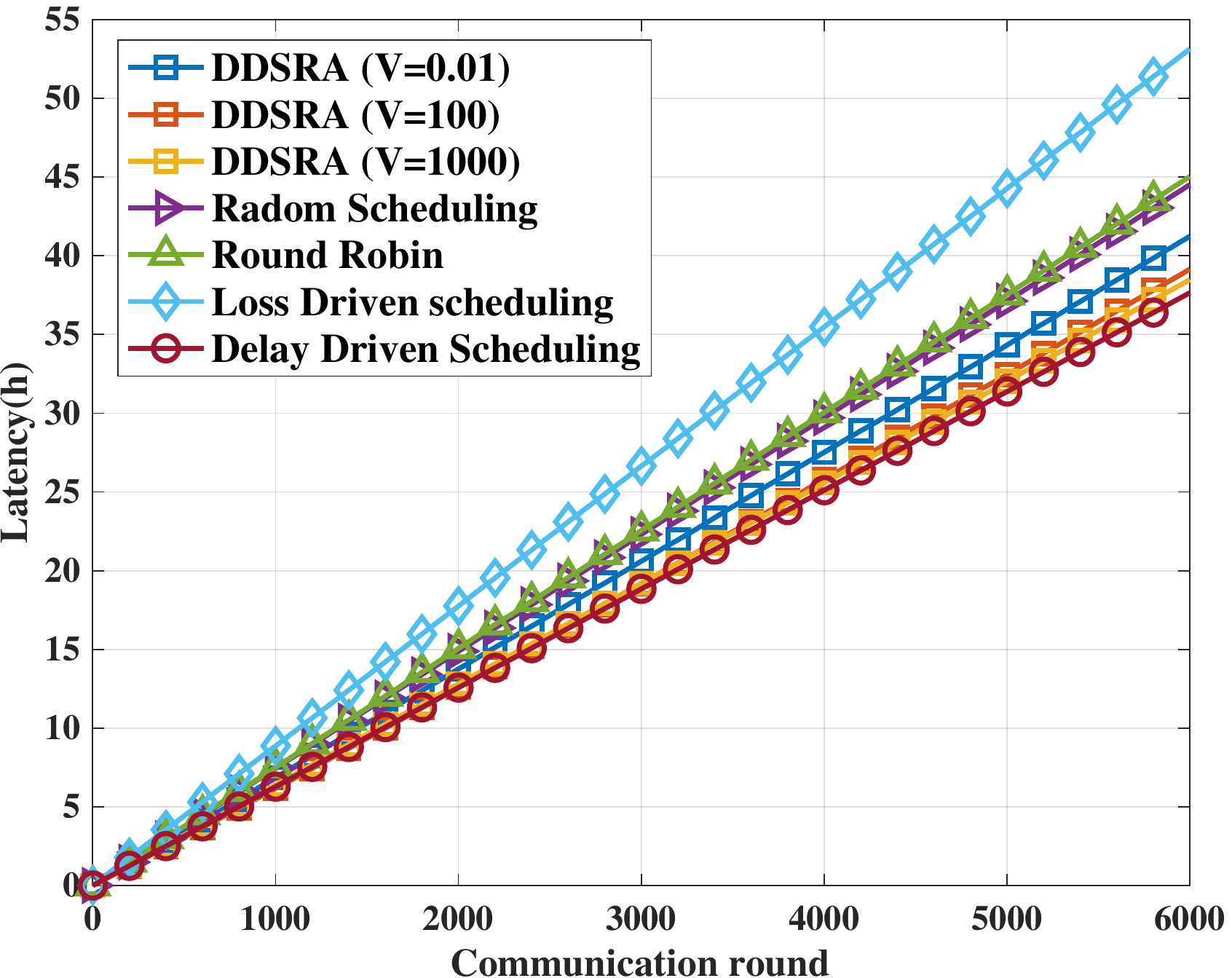}}
	\caption{Training delay comparison between DDSRA algorithm and the baseline schemes on (a) SVHN and (b) CIFAR-10 datasets.}
	\label{fig:14}
\end{figure}
Fig.\ref{fig:13} and \ref{fig:14} show the test accuracy and the training delay comparison between the proposed DDSRA algorithm (with $V=0.01, 1000$, and $10000$) and the baseline schemes, i.e., Random Scheduling, Round robin, Loss Driven Scheduling, and Delay Driven Scheduling. First, we can observe that a smaller $V$ can lead to a better FL performance but a higher FL training delay. It reveals that a smaller $V$ guaranteeing the proper participation rate of each gateway and associated devices can obtain a higher test accuracy while prolonging the training delay, which conforms to \textbf{Theorem} \ref{performance analysis}. Second, it can be observed that, with limited energy supply and memory, the proposed algorithm can achieve an obvious advantage on test accuracy and convergence rate than baseline schemes. Compared with Round Robin, the DDSRA algorithm with $V=0.01$ reduces the number of communication rounds required for convergence by $53\%$ for SVHN dataset and $78\%$ for CIFAR-10 dataset, and improves the test accuracy by $22\%$ for SVHN dataset and $37\%$ for CIFAR-10 dataset, respectively. The intuition is that the proposed DDSRA algorithm guarantees a proper participation rate of each gateway and associated devices, which makes the local datasets with better data distribution more involved in the FL training process. In addition, the joint communication, energy and memory resources allocation circumvents the local model training and transmitting failure due to the shortage of energy and memory, as such the gateways and devices can participate in more communication rounds to improve FL performance. Meanwhile, the baseline schemes fix the transmit power, computation frequency and the DNN partition point in the training process, as such devices and gateways often fail to complete the local model training and transmitting due to energy shortage. As a result, the low participation rate degrades the FL learning performance. Third, Fig.\ref{fig:14} shows that the proposed DDSRA algorithm achieves a much less FL latency than the baseline schemes, and the advantage of DDSRA algorithm is increasingly obvious as the communication round elapses. Compared with Loss Driven Scheduling, the DDSRA algorithm with $V=0.01$ reduces the training latency by $26\%$ for SVHN dataset and $23\%$ for CIFAR-10 dataset, respectively. Compared with Delay Driven Scheduling, the DDSRA algorithm with $V=0.01$ prolongs the training latency by $6\%$ for SVHN dataset and $7\%$ for CIFAR-10 dataset, while improving the test accuracy by $7\%$ for SVHN dataset and $17\%$ for CIFAR-10 dataset, respectively.

\begin{figure}[!t]
	\centering
	\subfigure[]{\includegraphics[width=3.2in]{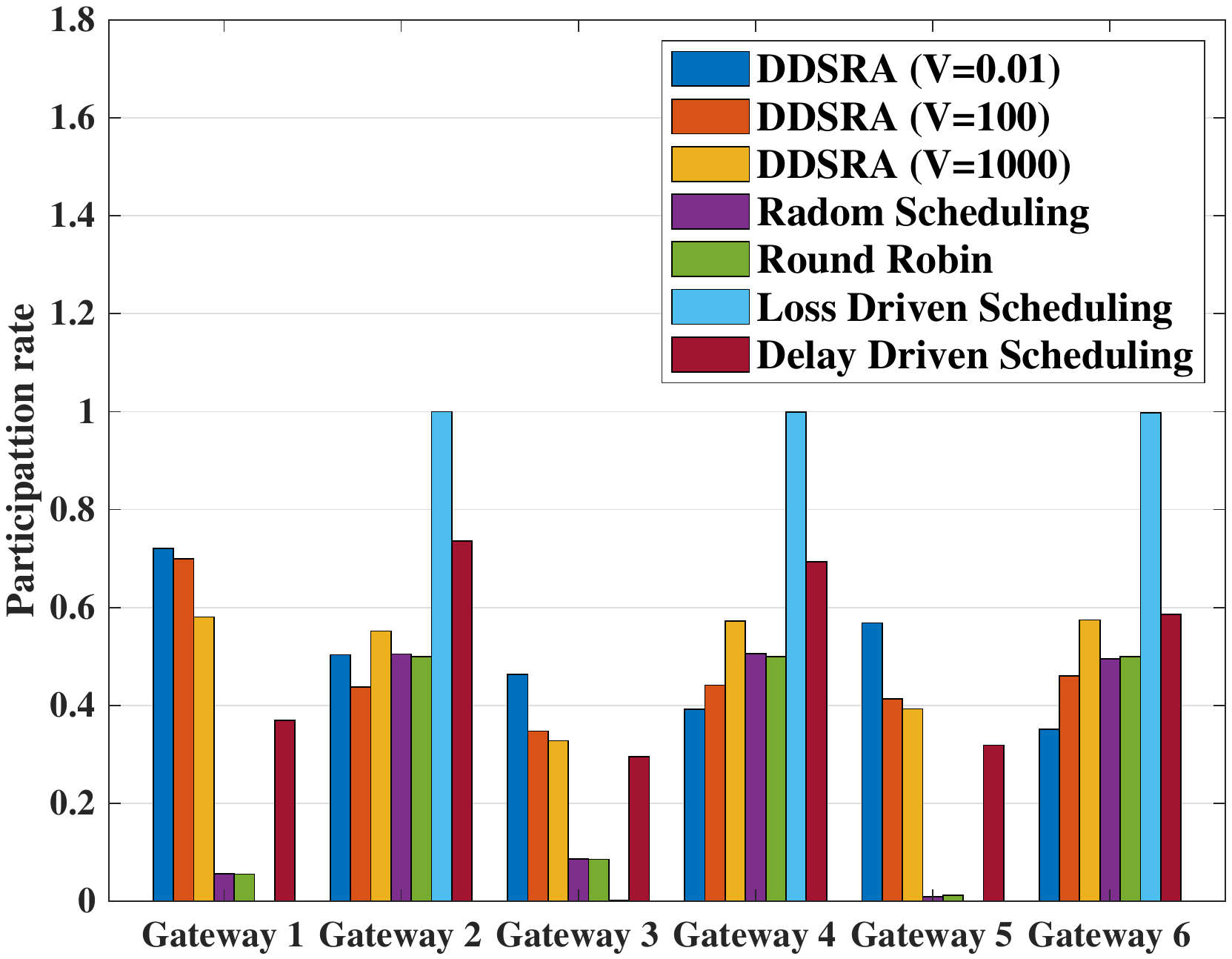}}
	\subfigure[]{\includegraphics[width=3.2in]{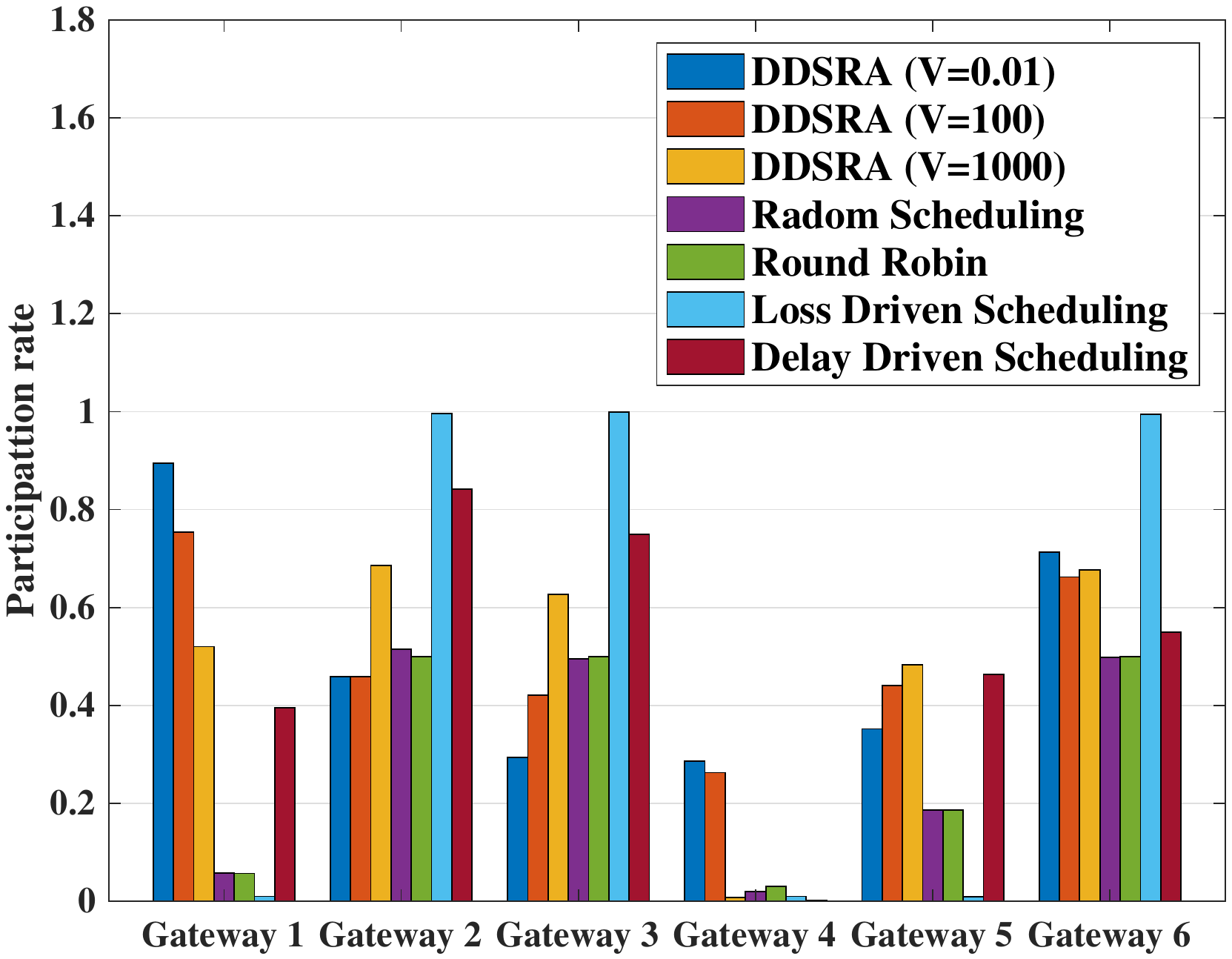}}
	\caption{Participation rate comparison between DDSRA algorithm and the baseline schemes on (a) SVHN and (b) CIFAR-10 datasets.}
	\label{fig:15}
\end{figure}
Fig.\ref{fig:15} shows the participation rate comparison between the proposed DDSRA algorithm (with $V=0.01, 1000$, and $10000$) and the baselines. First, it can be observed that for CIFAR-10 dataset, the $1$-th, $4$-th and $5$-th gateways and the respective associated devices rarely participate in FL training process in the Loss Driven Scheduling. This is due to that the local datasets at the devices associated with the $1$-th, $4$-th and $5$-th gateways are assigned with a wider variety of the non-IID data points than the other devices. That is, the $1$-th, $4$-th and $5$-th gateways and the respective associated devices are removed from the FL training process by the Loss Driven Scheduling since they achieve a lower training accuracy in each communication round. Second, Delay Driven Scheduling excludes the $4$-th gateway and its associated devices from the training process due to long transmission distance to the BS. This reduces the training latency at the cost of degrading the FL performance. Meanwhile, the proposed DDSRA algorithm saves the slow gateways and devices from being excluded from FL training process. To complete the FL training with limited harvested energy, the DDSRA algorithm lowers computational frequency and transmit power for the offloaded local model training and model transmitting, which improves the FL performance but increases the FL training latency. Third, the proposed algorithm achieves a much higher participation rate than the baselines, which contributes to the better learning performance as shown in Fig.\ref{fig:13}. In addition, it can be observed that a smaller $V$ encourages more gateways and devices participating in the FL process, which leads to a better FL performance. The experiential results shows that the proposed algorithm can not only save the slow devices from being excluded from FL training process, but also involve important devices in more communication rounds on the track of low latency by setting a larger participation rate for the important devices.

\section{Conclusion}\label{sec:Conclusion}
In this paper, we develop a communication-computation efficient FL framework for resource-limited IIoT networks that integrates DNN partition technique into the standard FL mechanism. By jointly optimizing the DNN partition point, channel assignment, transmit power, and computation frequency, the proposed DDSRA algorithm can be applied in a wide variety of device heterogeneity scenarios. With the developed device-specific participation rate, the DDSRA algorithm is robust against data heterogeneity by involving more devices with better data distribution over more communication rounds. Thanks to the layer-level memory usage and FLOPs calculation model, the DDSRA algorithm is widely applicable to other large-scale DNN models. Furthermore, we characterize a trade-off of [$\mathcal{O}(1/V)$, $\mathcal{O}(\sqrt{V})$] between the FL training latency minimization and the degree of which the participation rate constraint is satisfied with a control parameter $V$. The analytical convergence bound shows that the FL convergence rate can be improved by increasing the training data size and setting a higher participation rate for the important devices with better data distribution. Finally, experimental results demonstrate the developed device-specific participation rate in terms of feasibility. In addition, it has also been shown that DDSRA can obtain higher learning accuracy than the baselines under limited energy supply and memory capacity.

Several interesting directions immediately follow from this work. First, this work utilizes FLOPs to approximate the layer-level training latency and energy consumption. To provide a more accurate estimate, it is of interest to measure the latency and energy consumption by the DNN model training in real-world experiments. Second, due to the feedback loops in hidden layers, how to adapt the proposed FL framework to other large-scale artificial neural networks such as Recurrent Neural Network remains challenging.

\appendices
\section{Proof of Theorem \ref{theorem1}}
Before we show the main proof of \textbf{Theorem} \ref{theorem1}, we first give \textbf{Lemma} \ref{lemma1} below.
\begin{lemma}\label{lemma1}
	For any local epoch $k$ and communication round $t$, we have
	\begin{small}\begin{alignat}{1}\label{lemma1_1}
		\left\Vert \boldsymbol{w}_n^{k,t}-\boldsymbol{v}^{k,t}\right\Vert\leq \frac{\delta_n}{L_n}\left((\beta L_n+1)^k-1\right),
	\end{alignat}\end{small}and\begin{small}\begin{alignat}{1}\label{lemma1_2}
		\mathbb{E}{\left\Vert \tilde{\boldsymbol{w}}_n^{k,t}-\boldsymbol{w}_n^{k,t}\right\Vert}&\leq \frac{\sigma_n}{L_n\sqrt{\tilde{D}_n}}\left((\beta L_n+1)^k-1\right).
	\end{alignat}\end{small}
\end{lemma}
\begin{IEEEproof}
	The upper bound of $\left\Vert \boldsymbol{w}_n^{k,t}-\boldsymbol{v}^{k,t}\right\Vert$ in (\ref{lemma1_1}) is derived by induction. Initially, the upper bound of $\left\Vert \boldsymbol{w}_n^{k,t}-\boldsymbol{v}^{k,t}\right\Vert$ in (\ref{lemma1_1}) holds when $k=0$ since $ \boldsymbol{w}_n^{0,t}=\boldsymbol{v}^{0,t}$. Suppose that (\ref{lemma1_1}) holds at the $k$-th local epoch. Then, according to the update rule, it can be derived that\begin{small}\begin{alignat}{1}
		&\left\Vert \boldsymbol{w}_n^{k+1,t}-\boldsymbol{v}^{k+1,t}\right\Vert\!=\!\left\Vert \boldsymbol{w}_n^{k,t}\!-\!\beta\nabla F_n\!\left(\boldsymbol{w}_n^{k,t}\right)\!-\!\boldsymbol{v}^{k,t}\!+\!\beta\nabla F\!\left(\boldsymbol{v}^{k,t}\right)\right\Vert\nonumber\\&\leq\left\Vert \boldsymbol{w}_n^{k,t}-\boldsymbol{v}^{k,t}\right\Vert+\beta\left\Vert \nabla F_n\left(\boldsymbol{w}_n^{k,t}\right)-\nabla F_n\left(\boldsymbol{v}^{k,t}\right)\right\Vert+\beta\Big\Vert \nabla F_n\Big(\nonumber\\&\left.\left.\boldsymbol{v}^{k,t}\right)-\nabla F\left(\boldsymbol{v}^{k,t}\right)\right\Vert\leq(1+\beta L_n)\left\Vert \boldsymbol{w}_n^{k,t}-\boldsymbol{v}^{k,t}\right\Vert\!+\beta\delta_n\leq\frac{\delta_n}{L_n}\Big(\nonumber\\&(\beta L_n+1)^{k+1}-1\Big).
	\end{alignat}\end{small}As a result, the upper bound of $\left\Vert \boldsymbol{w}_n^{k,t}-\boldsymbol{v}^{k,t}\right\Vert$ in (\ref{lemma1_1}) also holds at the $(k+1)$-th local epoch. This concludes the proof of (\ref{lemma1_1}) in \textbf{Lemma} \ref{lemma1}.
	
	Note that from \textbf{Assumption} \ref{assumption3}, it can be derived that $\mathbb{E}\left\Vert \nabla \tilde{F}_n\left(\tilde{\boldsymbol{w}}_n^{k,t}\right)-\nabla F_n\left(\tilde{\boldsymbol{w}}_n^{k,t}\right)\right\Vert\leq\frac{\sigma_n}{\sqrt{\tilde{D}_n}}$. Similarly, the upper bound of $\mathbb{E}{\left\Vert \tilde{\boldsymbol{w}}_n^{k,t}-\boldsymbol{w}_n^{k,t}\right\Vert}$ can be obtained by induction.
\end{IEEEproof}

Based on \textbf{Lemma} \ref{lemma1}, it can be derived that\begin{small}\begin{alignat}{1}
	&\left\Vert\tilde{\boldsymbol{w}}_m^t-\boldsymbol{v}^{K,t}\right\Vert=\left\Vert\sum_n\frac{a_{m,n}\tilde{D}_n}{\sum_na_{m,n}\tilde{D}_n}\tilde{\boldsymbol{w}}_n^{K,t}-\boldsymbol{v}^{K,t}\right\Vert\nonumber\\&\leq\sum_n\frac{a_{m,n}\tilde{D}_n}{\sum_na_{m,n}\tilde{D}_n}\left(\left\Vert\tilde{\boldsymbol{w}}_n^{K,t}-\boldsymbol{w}_n^{K,t}\right\Vert+\left\Vert \boldsymbol{w}_n^{K,t}-\boldsymbol{v}^{K,t}\right\Vert\right)\nonumber\\&\leq\sum_n\frac{a_{m,n}\tilde{D}_n}{\sum_na_{m,n}\tilde{D}_n}\Big(\frac{\sigma_n}{L_n\sqrt{\tilde{D}_n}}+\frac{\delta_n}{L_n}\Big)\big((\beta L_n+1)^K-1\big).
\end{alignat}\end{small}This concludes the proof of \textbf{Theorem} \ref{theorem1}.

\section{Proof of Lemma \ref{bound of Deltat}}

First, from (\ref{vitual queues}), we have\begin{small}\begin{equation}\label{R1}
	Q_m(t+1)^2\leq Q_m(t)^2+(\Gamma_m-\mathbbm{1}_m^t)^2+2Q_m(t)(\Gamma_m-\mathbbm{1}_m^t).
\end{equation}\end{small}Next, by moving $Q_m(t)^2$ to the left-hand side of (\ref{R1}), dividing both sides by $2$, summing up the inequalities from $m=1$ to $M$, and taking the conditional expectation, it can be derived that \begin{small}\begin{alignat}{1}\label{R2}\Delta \Xi(t)&\leq \sum_{m\in \mathcal{M}}\mathbb{E}\left\{\left.Q_m(t)(\Gamma_m-\mathbbm{1}_m^t)\right\vert\boldsymbol{Q}(t)\right\}\nonumber\\&+\frac{1}{2}\sum_{m\in \mathcal{M}}\mathbb{E}\left\{\left.\Gamma_m+{\mathbbm{1}_m^t}^2\right\vert\boldsymbol{Q}(t)\right\}.
\end{alignat}\end{small}Note that $\mathbbm{1}_m^t=\sum_{j\in\mathcal{J}}I_{m,j}(t)$. Given constraints \textbf{C1} and \textbf{C3}, i.e., $I_{m,j}(t)\in \{0,1\},\forall m\in \mathcal{M},j\in \mathcal{J}$, and ${\sum}_{j\in\mathcal{J}}I_{m,j}(t)\leq 1, \forall j\in \mathcal{J}$, it can be derived that $0\leq{\mathbbm{1}_m^t}^2\leq 1$. Thus, the upper bound of the conditional Lyapunov drift $\Delta \Xi(t)$ can be derived as
\begin{small}\begin{equation}
	\Delta \Xi(t)\leq \frac{1}{2}\!\sum_{m\in \mathcal{M}}\!\!(\Gamma_m+1)+\!\!\!\sum_{m\in \mathcal{M}}\!\mathbb{E}\!\left\{\left.Q_m(t)(\Gamma_m \!-\!\mathbbm{1}_m^t)\right\vert\boldsymbol{Q}(t)\right\}.
\end{equation}\end{small}This concludes the proof of \textbf{Lemma} \ref{bound of Deltat}.

\section{Proof of Theorem \ref{performance analysis}}
Before we represent the main proof of \textbf{Theorem} \ref{performance analysis}, we first give \textbf{Lemma} \ref{lemma_appendixc} below.

\begin{lemma}\label{lemma_appendixc}
	For any $\varsigma>0$, there exists an IID policy $\pi'$ such that
	\begin{small}\begin{alignat}{1}
		\mathbb{E}\left\{\tau(t)\vert\pi'\right\}\leq \varphi^\text{opt}+\varsigma,\quad\mathbb{E}\{\mathbbm{1}_m^t\vert\pi'\}\ge \Gamma_m-\varsigma.\label{euq11:theorem}
	\end{alignat}\end{small}
\end{lemma}
\begin{IEEEproof}
	Given any $\varsigma>0$, we can note that there exists a policy $\pi^0$ which meets all of the constraints in \textbf{P0} and yields that $\lim_{T\to\infty}\inf \left[\frac{1}{T}\sum_{t=0}^{T-1}\mathbb{E}\{\tau(t)\vert\pi^0\}\right]\leq \varphi^\text{opt}+\varsigma$, and $\lim_{T\to\infty}\sup \left[\frac{1}{T}\sum_{t=0}^{T-1}\mathbb{E}\{\mathbbm{1}_m^t\vert\pi^0\}\right]\ge \Gamma_m-\varsigma$. For a integer $T_0$, it can be derived that
	\begin{small}\begin{alignat}{1}
		\frac{1}{T_0}{\sum}_{t=0}^{T_0-1}\mathbb{E}\{\tau(t)\vert\pi^0\}\leq \varphi^\text{opt}+\varsigma,\label{euq13:theorem}\\\frac{1}{T_0}{\sum}_{t=0}^{T_0-1}\mathbb{E}\{\mathbbm{1}_m^t\vert\pi^0\}\ge \Gamma_m-\varsigma.\label{euq9:theorem}
	\end{alignat}\end{small}From \cite{DBLP:journals/wc/NguyenDPSLNP21}, we can note that there exists an IID policy $\pi'$ such that
\begin{small}\begin{alignat}{1}\label{euq7:theorem}
\frac{1}{T_0}\sum_{t=0}^{T_0-1}\mathbb{E}\left\{\left.\left[\tau(t),\mathbbm{1}_1^t,...,\mathbbm{1}_M^t\right]\right\vert\pi^0\right\}=\mathbb{E}\left\{\left.\left[\tau(t),\mathbbm{1}_1^t,...,\mathbbm{1}_M^t\right]\right\vert\pi'\right\}.
\end{alignat}\end{small}Thus, by plugging (\ref{euq7:theorem}) into (\ref{euq13:theorem}) and (\ref{euq9:theorem}), we have (\ref{euq11:theorem}).
\end{IEEEproof}

Next, from \textbf{Lemma} \ref{bound of Deltat}, we have
\begin{small}\begin{alignat}{1}\label{euq12:theorem}
	\Delta_V(t)\!\leq\! H\!+\!\!\!\sum_{m\in \mathcal{M}}\mathbb{E}\left\{V\tau(t)+Q_m(t)(\Gamma_m\!-\!\mathbbm{1}_m^t)\vert\boldsymbol{Q}(t),\pi'\right\}.
\end{alignat}\end{small}Plugging (\ref{euq11:theorem}) into the right-hand-side of (\ref{euq12:theorem}), letting $\varsigma\to 0$, and taking expectation of both sides, we have\begin{small}\begin{alignat}{1}\label{equ4:theorem1}
	\mathbb{E}\{\Xi(t+1)-\Xi(t)\vert\boldsymbol{Q}(t)\}+V\mathbb{E}\{\tau(t)\vert\boldsymbol{Q}(t)\}\leq H+V\varphi^\text{opt}.
\end{alignat}\end{small}By summing up (\ref{equ4:theorem1}) form $t=0$ to $T-1$, and dividing both sides by $T$ and $V$, we have \begin{small}\begin{alignat}{1}\frac{\sum_{t=1}^T\tau(t)}{T}\leq \varphi^\text{opt}+\frac{H}{V}+\frac{\mathbb{E}\{\Xi(0)-\Xi(T)\}}{VT},
\end{alignat}\end{small}which concludes the proof of (\ref{equ2:theorem1}).

Next, from (\ref{equ4:theorem1}), it can be derived that \begin{small}\begin{alignat}{1}\label{equ55}\Delta\Xi(t))\leq H+V(\varphi^\text{opt}-\tau^\text{min}),
\end{alignat}\end{small}where $\tau^\text{min}=\frac{K\min_{n\in\mathcal{N}}\left\{\tilde{D}_n\right\}\sum_{l=1}^L(o_l+o'_l)}{\min\{\min_{n\in\mathcal{N}}\{\phi_n^\text{D}f_n^\text{D}\},\min_{m\in\mathcal{M}}\{\phi_m^\text{G}f_m^\text{G,max}\}\}}+{\gamma}/B^\text{u}/$ $\log_2\left(1+\frac{P_m^\text{max}\overline{h^u_{m,j}}}{(B^\text{u}N_0+\overline{I^u_{m,j}})}\right)+{\gamma}/{B^\text{d}/\log_2\left(1+\frac{P^B\overline{h^d_{m,j}}}{(B^\text{d}N_0+\overline{I^d_{m,j}})}\right)}$. By summing up (\ref{equ55}) from $t=0$ to $T-1$, taking expectations, dividing both sides by $T$, and recalling that $\Xi(t)=\frac{1}{2}\sum_{m\in \mathcal{M}}Q_m(t)^2$, it can be derived that\begin{small}\begin{alignat}{1}\sum_{m\in \mathcal{M}}\!\!\frac{\mathbb{E}\{Q_m(T)^2\}}{T}\leq H \!+\! V(\varphi^\text{opt}-\tau^\text{min})\!+\!\!\!\sum_{m\in \mathcal{M}}\!\!\frac{\mathbb{E}\{Q_m(0)^2\}}{T}.
\end{alignat}\end{small}Thus, for each gateway and the associated devices, we have\begin{small}\begin{alignat}{1}\label{equ:108}
	\frac{\mathbb{E}\{Q_m(T)^2\}}{T}\leq H+V(\varphi^\text{opt}-\tau^\text{min})+\sum_{m\in \mathcal{M}}\frac{\mathbb{E}\{Q_m(0)^2\}}{T}.
\end{alignat}\end{small}By dividing both sides of (\ref{equ:108}) by $T$, and taking the square root of both sides, we have\begin{small}\begin{alignat}{1}\label{equ19:theorem}
	\frac{\mathbb{E}\{Q_m(T)\}}{T}\leq\sqrt{ \frac{H+V(\varphi^\text{opt}-\tau^\text{min})}{T}+\sum_{m\in \mathcal{M}}\!\!\frac{\mathbb{E}\{Q_m(0)^2\}}{T^2}}.
\end{alignat}\end{small}From (\ref{vitual queues}), it can be derived that\begin{small}\begin{alignat}{1}\label{equ17:theorem1}
	Q_m(t+1)\ge Q_m(t)-\mathbbm{1}_m^t+\Gamma_m.
\end{alignat}\end{small}Note that $\mathbb{E}\{Q_m(0)\}<\infty$. By summing up (\ref{equ17:theorem1}) from $t=0$ to $T-1$, taking expectations, and dividing both sides by $T$, it can be derived that $\frac{\mathbb{E}\{Q_m(T)\}}{T}\ge \Gamma_m-\frac{1}{T}{\sum}_{t=0}^{T-1}\mathbbm{1}_m^t$. Thus, from (\ref{equ19:theorem}), we have\begin{small}\begin{alignat}{1}
		\frac{1}{T}\sum_{t=0}^{T-1}\mathbbm{1}_m^t\ge \Gamma_m\!-\!\!\sqrt{ \frac{H\!+\!V(\varphi^\text{opt}\!-\!\tau^\text{min})}{T}+\!\!\!\sum_{m\in \mathcal{M}}\!\!\!\frac{\mathbb{E}\{Q_m(0)^2\}}{T^2}},
\end{alignat}\end{small}This concludes the proof of (\ref{equ3:theorem1}).

\section{Proof of Theorem \ref{theorem2}}
Before we show the main proof of \textbf{Theorem} \ref{theorem2}, we first give \textbf{Lemma} \ref{lemma2} below.
\begin{lemma}\label{lemma2}
	For any local epoch $k$ and communication round $t$, we have
	\begin{small}\begin{alignat}{1}\label{lemma2_1}
		\left\Vert \boldsymbol{w}^{k,t}-\boldsymbol{v}^{k,t}\right\Vert\leq \frac{\delta}{L}\left((\beta L+1)^k-1\right)-\beta\delta k,
	\end{alignat}\end{small}\begin{small}\begin{alignat}{1}\label{lemma2_2}
		\mathbb{E}\left\Vert \tilde{\boldsymbol{w}}^{k,t}-\boldsymbol{w}^{k,t}\right\Vert&\leq\frac{1}{L}\left(\sum_{n\in\mathcal{N}}\xi_n\frac{\sigma_n}{\sqrt{D_n(t)}}\right)\left((\beta L+1)^k-1\right)\nonumber\\&+\beta k\left(\sum_{n\in\mathcal{N}}\left\vert\xi_n-\frac{D_n}{\sum_{n\in\mathcal{N}}D_n}\right\vert\rho_n\right),
	\end{alignat}\end{small}where $\xi_n=\frac{\sum_{m\in\mathcal{M}} \Gamma_ma_{m,n}\tilde{D}_n}{\sum_{n\in\mathcal{N}}\sum_{m\in\mathcal{M}} \Gamma_ma_{m,n}\tilde{D}_n}$.
\end{lemma}

\begin{IEEEproof}
	In this proof, we first derive the upper bound of $\left\Vert \boldsymbol{w}^{k,t}-\boldsymbol{v}^{k,t}\right\Vert$ in (\ref{lemma2_1}) by induction. Initially, the upper bound of $\left\Vert \boldsymbol{w}^{k,t}-\boldsymbol{v}^{k,t}\right\Vert$ in (\ref{lemma2_1}) holds at $k=0$ since $ \boldsymbol{w}^{0,t}=\boldsymbol{v}^{0,t}$. Suppose that (\ref{lemma2_1}) holds at the $k$-th local epoch. According to the update rule, it can be derived that\begin{small}\begin{alignat}{1}
		&\left\Vert\boldsymbol{w}^{k+1,t}\!-\!\boldsymbol{v}^{k+1,t}\right\Vert\!=\!\left\Vert\boldsymbol{w}^{k,t}\!-\!\boldsymbol{v}^{k,t}\!-\!\frac{\beta}{\sum_n\!D_n}{\sum}_n\!D_n\left(\nabla F_n\!\left(\boldsymbol{w}_n^{k,t}\right)\right.\right.\nonumber\\&\left.\left.-\nabla F_n\!\left(\boldsymbol{v}^{k,t}\right)\right)\right\Vert\!\leq\!\left\Vert\boldsymbol{w}^{k,t}\!-\!\boldsymbol{v}^{k,t}\right\Vert\!+\!\frac{\beta}{\sum_n\!D_n}{\sum}_n\!D_n\left\Vert\nabla F_n\!\left(\boldsymbol{w}_n^{k,t}\right)\right.\nonumber\\&\left.-\nabla F_n\!\left(\boldsymbol{v}^{k,t}\right)\right\Vert\!\leq\!\left\Vert\boldsymbol{w}^{k,t}-\boldsymbol{v}^{k,t}\right\Vert\!+\!\frac{\beta L}{\sum_n\!D_n}{\sum}_n\!D_n\left\Vert \boldsymbol{w}_n^{k,t}-\boldsymbol{v}^{k,t}\right\Vert\nonumber\\&\leq\left\Vert\boldsymbol{w}^{k,t}-\boldsymbol{v}^{k,t}\right\Vert+\beta\delta(\beta L+1)^k-\beta\delta\leq\frac{\delta}{L}\left((\beta L+1)^{k+1}-1\right)\nonumber\\&-\beta\delta(k+1).
\end{alignat}\end{small}As a result, the upper bound of $\left\Vert\boldsymbol{w}^{k,t}-\boldsymbol{v}^{k,t}\right\Vert$ in (\ref{lemma2_1}) also holds at the $(k+1)$-th local epoch. This concludes the proof of (\ref{lemma2_1}) in \textbf{Lemma} \ref{lemma2}.

Next, we obtain the upper bound of $\mathbb{E}{\left\Vert \tilde{\boldsymbol{w}}^{k,t}-\boldsymbol{w}^{k,t}\right\Vert}$ by induction as follows. Initially, the upper bound of $\mathbb{E}{\left\Vert \tilde{\boldsymbol{w}}_n^{k,t}-\boldsymbol{w}_n^{k,t}\right\Vert}$ holds at $k=0$ since $ \tilde{\boldsymbol{w}}_n^{0,t}=\boldsymbol{w}_n^{0,t}$. Suppose that the upper bound of $\mathbb{E}{\left\Vert \tilde{\boldsymbol{w}}_n^{k,t}-\boldsymbol{w}_n^{k,t}\right\Vert}$ in (\ref{lemma1_2}) holds at the $k$-th local epoch. Thus, it can be derived that\begin{small}\begin{alignat}{1}
		&\left\Vert \tilde{\boldsymbol{w}}^{k+1,t}-\boldsymbol{w}^{k+1,t}\right\Vert=\left\Vert \tilde{\boldsymbol{w}}^{k,t}-\beta\sum_n\!\frac{\sum_m \mathbbm{1}_m^ta_{m,n}\tilde{D}_n}{\sum_n\sum_m \mathbbm{1}_m^ta_{m,n}\tilde{D}_n}\nabla \tilde{F}_n\!\Big(\right.\nonumber\\&\left.\left.\tilde{\boldsymbol{w}}_n^{k,t}\right)\!-\!\boldsymbol{w}^{k,t}\!+\!\beta\sum_n\frac{D_n}{\sum_nD_n}\nabla F_n\left(\boldsymbol{w}_n^{k,t}\right)\right\Vert\!\leq\!\left\Vert\tilde{\boldsymbol{w}}^{k,t}-\boldsymbol{w}^{k,t}\right\Vert\!+\!\beta\nonumber\\&\left\Vert \sum_n\!\frac{\sum_m \mathbbm{1}_m^ta_{m,n}\tilde{D}_n}{\sum_n\!\sum_m\! \mathbbm{1}_m^ta_{m,n}\tilde{D}_n}\nabla \tilde{F}_n\!\!\left(\tilde{\boldsymbol{w}}_n^{k,t}\right)\!-\!\sum_n\frac{D_n}{\sum_n\!D_n}\nabla F_n\!\left(\boldsymbol{w}_n^{k,t}\right)\right\Vert\nonumber\\&\leq\left\Vert \tilde{\boldsymbol{w}}^{k,t}-\boldsymbol{w}^{k,t}\right\Vert+\beta\sum_n\Bigg\Vert\left(\nabla \tilde{F}_n\left(\tilde{\boldsymbol{w}}_n^{k,t}\right)\!-\!\nabla F_n\!\left(\tilde{\boldsymbol{w}}_n^{k,t}\right)\!+\!\nabla\! F_n\!\Big(\right.\nonumber\\&\left.\left.\left.\tilde{\boldsymbol{w}}_n^{k,t}\right)\!-\!\nabla\! F_n\!\left(\boldsymbol{w}_n^{k,t}\right)\right)\!\frac{\sum_m \mathbbm{1}_m^ta_{m,n}\tilde{D}_n}{\sum_n\!\sum_m \mathbbm{1}_m^ta_{m,n}\tilde{D}_n}\right\Vert\!+\!\beta\!\sum_n\!\left\Vert\nabla\! F_n\!\left(\boldsymbol{w}_n^{k,t}\right)\right.\nonumber\\&\left.\left(\frac{\sum_m \mathbbm{1}_m^ta_{m,n}\tilde{D}_n}{\sum_n\sum_m \mathbbm{1}_m^ta_{m,n}\tilde{D}_n}-\frac{D_n}{\sum_nD_n}\right)\right\Vert\leq\left\Vert \tilde{\boldsymbol{w}}^{k,t}\!-\!\boldsymbol{w}^{k,t}\right\Vert\!+\!\beta\sum_n\Bigg\Vert\nonumber\\&\frac{\sum_m \mathbbm{1}_m^ta_{m,n}\tilde{D}_n}{\sum_n\!\sum_m\! \mathbbm{1}_m^ta_{m,n}\tilde{D}_n}\left(\nabla\! \tilde{F}_n\!\left(\tilde{\boldsymbol{w}}_n^{k,t}\right)\!-\!\nabla\! F_n\!\left(\tilde{\boldsymbol{w}}_n^{k,t}\right)\right)\Bigg\Vert\!+\!\beta\Bigg\Vert\Big(\nabla F_n\Big(\nonumber\\&\left.\left.\left.\tilde{\boldsymbol{w}}_n^{k,t}\right)\!-\!\nabla F_n\!\!\left(\boldsymbol{w}_n^{k,t}\right)\right)\!\frac{\sum_m \mathbbm{1}_m^ta_{m,n}\tilde{D}_n}{\sum_n\!\sum_m\! \mathbbm{1}_m^ta_{m,n}\tilde{D}_n}\right\Vert\!+\!\beta\!\sum_n\!\Bigg\Vert\nabla F_n\!\left(\!\boldsymbol{w}_n^{k,t}\right)\nonumber\\&\left(\frac{\sum_m \mathbbm{1}_m^ta_{m,n}\tilde{D}_n}{\sum_n\sum_m \mathbbm{1}_m^ta_{m,n}\tilde{D}_n}-\frac{D_n}{\sum_nD_n}\right)\Bigg\Vert\leq\left\Vert \tilde{\boldsymbol{w}}^{k,t}-\boldsymbol{w}^{k,t}\right\Vert+\beta\sum_n\nonumber\\&\left\Vert\frac{\sum_m \mathbbm{1}_m^ta_{m,n}\tilde{D}_n}{\sum_n\sum_m \mathbbm{1}_m^ta_{m,n}\tilde{D}_n}\right\Vert\left(\left\Vert\nabla \tilde{F}_n\left(\tilde{\boldsymbol{w}}_n^{k,t}\right)\!-\!\nabla F_n\left(\tilde{\boldsymbol{w}}_n^{k,t}\right)\right\Vert\!+\!\Big\Vert\nabla F_n\right.\nonumber\\&\!\left.\left.\left(\tilde{\boldsymbol{w}}_n^{k,t}\right)\!-\!\nabla F_n\!\left(\!\boldsymbol{w}_n^{k,t}\right)\right\Vert\right)\!+\!\beta\!\sum_n\!\Bigg\Vert\frac{\sum_m \mathbbm{1}_m^ta_{m,n}\tilde{D}_n}{\sum_n\!\sum_m\! \mathbbm{1}_m^ta_{m,n}\tilde{D}_n}\!-\!\frac{D_n}{\sum_n\!\!D_n}\Bigg\Vert\nonumber\\&\left\Vert\nabla F_n\left(\boldsymbol{w}_n^{k,t}\right)\right\Vert.
\end{alignat}\end{small}Based on (\ref{lemma1_2}) in \textbf{Lemma} \ref{lemma1} and \textbf{Assumption} \ref{assumption3}, we have\begin{small}\begin{alignat}{1}
		&\left\Vert \tilde{\boldsymbol{w}}^{k+1,t}-\boldsymbol{w}^{k+1,t}\right\Vert\leq\left\Vert \tilde{\boldsymbol{w}}^{k,t}-\boldsymbol{w}^{k,t}\right\Vert\!+\!\beta\sum_n\Bigg(\frac{\sigma_n}{\sqrt{\tilde{D}_n}}\!+\!\frac{\sigma_n}{\sqrt{\tilde{D}_n}}\Big(\nonumber\\&(\beta L_n+1)^k-1\Big)\Bigg)\left\Vert\frac{\sum_m \mathbbm{1}_m^ta_{m,n}\tilde{D}_n}{\sum_n\sum_m \mathbbm{1}_m^ta_{m,n}\tilde{D}_n}\right\Vert\!+\!\beta\sum_n\left\Vert-\frac{D_n}{\sum_nD_n}+\right.\nonumber\\&\left.\frac{\sum_m \mathbbm{1}_m^ta_{m,n}\tilde{D}_n}{\sum_n\sum_m \mathbbm{1}_m^ta_{m,n}\tilde{D}_n}\right\Vert\rho_n.
\end{alignat}\end{small}By taking expectation of both sides, we have \begin{small}\begin{alignat}{1}\label{lemma2_3}
		&\mathbb{E}\left\Vert \tilde{\boldsymbol{w}}^{k+1,t}-\boldsymbol{w}^{k+1,t}\right\Vert\leq\mathbb{E}\left\Vert \tilde{\boldsymbol{w}}^{k,t}-\boldsymbol{w}^{k,t}\right\Vert+\beta{\sum}_n\left\Vert\xi_n\right\Vert\frac{\sigma_n}{\sqrt{\tilde{D}_n}}\nonumber\\&(\beta L_n+1)^k+\beta{\sum}_n\left\Vert\xi_n-\frac{D_n}{{\sum}_nD_n}\right\Vert\rho_n.\end{alignat}\end{small}Plugging (\ref{lemma2_2}) in \textbf{Lemma} \ref{lemma2} into the right-hand-side of (\ref{lemma2_3}), it can be proved that the upper bound of $\mathbb{E}{\left\Vert \tilde{\boldsymbol{w}}^{k,t}-\boldsymbol{w}^{k,t}\right\Vert}$ in (\ref{lemma2_2}) also holds at the $(k+1)$-th local epoch. This concludes the proof of (\ref{lemma2_2}) in \textbf{Lemma} \ref{lemma2}.
\end{IEEEproof}

Thus, based on \textbf{Lemma} \ref{lemma2}, it can be derived that\begin{small}\begin{alignat}{1}\label{theorem2_2}
	\mathbb{E}\left\Vert\tilde{\boldsymbol{w}}^{k,t}-\boldsymbol{v}^{k,t}\right\Vert&\leq\frac{1}{L}\left(\delta+{\sum}_n\xi_n\frac{\sigma_n}{\sqrt{\tilde{D}_n}}\right)\left((\beta L+1)^k-1\right)\nonumber\\&+\beta k\left(\delta+{\sum}_n\left\vert\xi_n-\frac{D_n}{{\sum}_nD_n}\right\vert\rho_n\right).
\end{alignat}\end{small}Based on (\ref{theorem2_2}), the detailed proof of \textbf{Theorem} \ref{theorem2} can be found in \cite{DBLP:journals/jsac/WangTSLMHC19}.

\section{Proof of Theorem \ref{theorem3}}
First, due to $F(\boldsymbol{w})$ is $L$-smooth, it can be derived that\begin{small}\begin{alignat}{1}\label{theorem3_2}
	&F\left(\tilde{\boldsymbol{w}}^{t+1}\right)-F\left(\tilde{\boldsymbol{w}}^t\right)\nonumber\\&\leq \frac{L}{2}\left\Vert\tilde{\boldsymbol{w}}^{t+1}-\tilde{\boldsymbol{w}}^t\right\Vert^2+\left\langle\nabla F\left(\tilde{\boldsymbol{w}}^t\right),\tilde{\boldsymbol{w}}^{t+1}-\tilde{\boldsymbol{w}}^t\right\rangle.
\end{alignat}\end{small}Taking the conditional expectation on both sides of (\ref{theorem3_2}), we have\begin{small}\begin{alignat}{1}
	&\mathbb{E}\left[\left.F\left(\tilde{\boldsymbol{w}}^{t+1}\right)\right\vert\tilde{\boldsymbol{w}}^t\right]-F\left(\tilde{\boldsymbol{w}}^t\right)\leq\frac{L}{2}\mathbb{E}\left[\left.\left\Vert\tilde{\boldsymbol{w}}^{t+1}-\tilde{\boldsymbol{w}}^t\right\Vert^2\right\vert\tilde{\boldsymbol{w}}^t\right]\nonumber\\&+\left\langle\nabla F\left(\tilde{\boldsymbol{w}}^t\right),\mathbb{E}\left[\left.\tilde{\boldsymbol{w}}^{t+1}-\tilde{\boldsymbol{w}}^t\right\vert\tilde{\boldsymbol{w}}^t\right]\right\rangle.
\end{alignat}\end{small}Second, according to the update rule, we have\begin{small}\begin{alignat}{1}\label{theorem3_3}
	&\mathbb{E}\left[\left.F\left(\tilde{\boldsymbol{w}}^{t+1}\right)\right\vert\tilde{\boldsymbol{w}}^t\right]\!-\!F\left(\tilde{\boldsymbol{w}}^t\right)=\frac{L\beta^2}{2}\mathbb{E}\left[\left\Vert\sum_n\frac{\sum_m \mathbbm{1}_m^ta_{m,n}\tilde{D}_n}{\sum_n\sum_m \mathbbm{1}_m^ta_{m,n}\tilde{D}_n}\right.\right.\nonumber\\&\left.\left.\sum_{k=0}^{K-1}\!\nabla F_n\!\!\left(\tilde{\boldsymbol{w}}_n^{k,t}\right)\right\Vert^2\Bigg\vert\tilde{\boldsymbol{w}}^t\right]\!+\!\left\langle\!\nabla\! F\left(\tilde{\boldsymbol{w}}^t\right),\mathbb{E}\!\left[-\beta \sum_n\!\sum_{k=0}^{K-1}\!\nabla\! F_n\!\!\left(\tilde{\boldsymbol{w}}_n^{k,t}\right)\right.\right.\nonumber\\&\left.\left.\frac{\sum_m \mathbbm{1}_m^ta_{m,n}\tilde{D}_n}{\sum_n\!\sum_m\! \mathbbm{1}_m^ta_{m,n}\tilde{D}_n} \Bigg\vert\tilde{\boldsymbol{w}}^t\right]\!\right\rangle\leq \frac{L\beta^2NK}{2}\mathbb{E}\left[\sum_n\!\sum_{k=0}^{K-1}\left\Vert\nabla\! F_n\!\left(\tilde{\boldsymbol{w}}_n^{k,t}\right)\right.\right.\nonumber\\&\left.\left.\frac{\sum_m\! \mathbbm{1}_m^ta_{m,n}\tilde{D}_n}{\sum_n\!\sum_m \!\!\mathbbm{1}_m^ta_{m,n}\tilde{D}_n}\right\Vert^2\!\Bigg\vert\tilde{\boldsymbol{w}}^t\!\right]\!\!+\!\left\langle\!\mathbb{E}\!\left[\!-\beta\! \sum_n\!\frac{\sum_m \mathbbm{1}_m^ta_{m,n}\tilde{D}_n}{\sum_n\!\sum_m\! \mathbbm{1}_m^ta_{m,n}\tilde{D}_n}\!\!\sum_{k=0}^{K-1}\right.\right.\nonumber\\&\left.\left.\nabla\! F_n\!\!\left(\!\tilde{\boldsymbol{w}}_n^{k,t}\!\right) \Bigg\vert\tilde{\boldsymbol{w}}^t\right]\!,\nabla\! F\!\left(\tilde{\boldsymbol{w}}^t\right)\!\right\rangle\!=\! \frac{L\beta^2NK}{2}\!\!\sum_n\!\sum_{k=0}^{K-1}\!\xi_n^2\mathbb{E}\!\bigg[\left\Vert\nabla\! F_n\!\left(\!\tilde{\boldsymbol{w}}_n^{k,t}\!\right)\right.\nonumber\\&\;\;\Big\Vert^2\bigg\vert\tilde{\boldsymbol{w}}^t\bigg]\!+\!\left\langle\!\nabla F\!\left(\tilde{\boldsymbol{w}}^t\right),-\beta\sum_n\!\sum_{k=0}^{K-1}\xi_n\mathbb{E}\left[\nabla F_n\!\left(\!\tilde{\boldsymbol{w}}_n^{k,t}\!\right) \Big\vert\tilde{\boldsymbol{w}}^t\right]\right\rangle.\!\!\!
\end{alignat}\end{small}Note that $\xi_n=\frac{\sum_{m\in\mathcal{M}} \Gamma_ma_{m,n}\tilde{D}_n}{\sum_{n\in\mathcal{N}}\sum_{m\in\mathcal{M}} \Gamma_ma_{m,n}\tilde{D}_n}$. Taking the expectation on both sides of (\ref{theorem3_3}), we have\begin{small}\begin{alignat}{1}\label{theorem3_5}
	&\mathbb{E}\left[F\!\left(\tilde{\boldsymbol{w}}^{t+1}\right)\right]\!-\!\mathbb{E}\left[F\!\left(\tilde{\boldsymbol{w}}^t\right)\right]\leq\frac{L\beta^2NK}{2}\!\sum_n\!\sum_{k=0}^{K-1}\xi_n^2\mathbb{E}\bigg[\Big\Vert\nabla F_n\!\left(\tilde{\boldsymbol{w}}_n^{k,t}\right)\nonumber\\&\Big\Vert^2\bigg]\!+\!\left\langle\mathbb{E}\left[\nabla F\!\left(\tilde{\boldsymbol{w}}^t\right)\right],-\beta \sum_n\!\sum_{k=0}^{K-1}\xi_n\mathbb{E}\!\left[\nabla F_n\!\left(\tilde{\boldsymbol{w}}_n^{k,t}\right)\right]\!\right\rangle\!=\!\frac{L\beta^2NK}{2}\nonumber\\&\sum_n\!\sum_{k=0}^{K-1}\xi_n^2\mathbb{E}\!\left[\left\Vert\nabla F_n\!\left(\tilde{\boldsymbol{w}}_n^{k,t}\right)\right\Vert^2\right]\!+\!\beta\!\sum_{k=0}^{K-1}\!\mathbb{E}\Bigg[\Bigg\langle\!-\!\sum_n\xi_n\nabla F_n\!\left(\tilde{\boldsymbol{w}}_n^{k,t}\right),\nonumber\\&\nabla F\left(\tilde{\boldsymbol{w}}^t\right)\Bigg\rangle\Bigg].
\end{alignat}\end{small}Note that\begin{small}\begin{alignat}{1}\label{theorem3_4}
	&\mathbb{E}\left[\left\langle\!\nabla F\!\left(\tilde{\boldsymbol{w}}^t\right),-\!\sum_n\xi_n\nabla F_n\!\left(\tilde{\boldsymbol{w}}_n^{k,t}\right)\!\right\rangle\right]\!=\!\mathbb{E}\Bigg[\Bigg\langle\nabla F\!\left(\tilde{\boldsymbol{w}}^t\right),\nabla F\!\left(\tilde{\boldsymbol{w}}^t\right)\nonumber\\&-\!\nabla F\!\left(\tilde{\boldsymbol{w}}^t\right)\!-\!\sum_n\xi_n\nabla F_n\!\left(\tilde{\boldsymbol{w}}_n^{k,t}\right)\!\Bigg\rangle\Bigg]\!=\!-\mathbb{E}\left[\left\langle\nabla F\!\left(\tilde{\boldsymbol{w}}^t\right),\nabla F\!\left(\tilde{\boldsymbol{w}}^t\right)\right\rangle\right]\nonumber\\&+\mathbb{E}\left[\left\langle\nabla F\left(\tilde{\boldsymbol{w}}^t\right),\nabla F\left(\tilde{\boldsymbol{w}}^t\right)- \sum_n\xi_n\nabla F_n\left(\tilde{\boldsymbol{w}}_n^{k,t}\right)\right\rangle\right]\leq-\frac{1}{2}\mathbb{E}\Big[\big\Vert\nonumber\\&\nabla F\!\left(\tilde{\boldsymbol{w}}^t\right)\big\Vert^2\Big]\!+\!\frac{1}{2}\mathbb{E}\left[\left\Vert \nabla F\!\left(\tilde{\boldsymbol{w}}^t\right)\!-\!\sum_n\xi_n\nabla F_n\left(\!\tilde{\boldsymbol{w}}_n^{k,t}\!\right)\right\Vert^2\right]\!\leq-\frac{1}{2}\mathbb{E}\Big[\big\Vert\nonumber\\&\left.\left.\!\! \nabla F\left(\tilde{\boldsymbol{w}}^t\right)\right\Vert^2\right]+\frac{N}{2} \sum_n\xi_n^2\mathbb{E}\left[\left\Vert\nabla F\left(\tilde{\boldsymbol{w}}^t\right)-\nabla F_n\left(\!\tilde{\boldsymbol{w}}_n^{k,t}\!\right)\right\Vert^2\right]\leq-\frac{1}{2}\nonumber\\&\mathbb{E}\left[\left\Vert \nabla F\left(\tilde{\boldsymbol{w}}^t\right)\right\Vert^2\right]\!+\!\frac{N}{2}\! \sum_n\xi_n^2L_n^2\mathbb{E}\left[\left\Vert\tilde{\boldsymbol{w}}^t\!-\!\tilde{\boldsymbol{w}}_n^{k,t}\right\Vert^2\right]\leq\!-\!\frac{1}{2}\mathbb{E}\Big[\big\Vert \nabla F\big(\nonumber\\&\left.\tilde{\boldsymbol{w}}^t\right)\big\Vert^2\Big]\!+\!\frac{N}{2} \sum_n\xi_n^2L_n^2\beta^2\mathbb{E}\left[\left\Vert\sum_{j=0}^{k-1}\nabla F_n\!\left(\tilde{\boldsymbol{w}}_n^{j,t}\right)\right\Vert^2\right]\!\leq\!-\frac{1}{2}\mathbb{E}\Big[\big\Vert \nabla F\nonumber\\&\left(\tilde{\boldsymbol{w}}^t\right)\big\Vert^2\Big]+\frac{N}{2} \sum_n\xi_n^2L_n^2\beta^2k\sum_{j=0}^{k-1}\mathbb{E}\left[\left\Vert\nabla F_n\left(\tilde{\boldsymbol{w}}_n^{j,t}\right)\right\Vert^2\right].
\end{alignat}\end{small}Plugging (\ref{theorem3_4}) into the right-hand-side of (\ref{theorem3_5}), we have\begin{small}\begin{alignat}{1}\label{theorem3_6}
	&\mathbb{E}\left[F\left(\tilde{\boldsymbol{w}}^{t+1}\right)\right]-\mathbb{E}\left[F\left(\tilde{\boldsymbol{w}}^t\right)\right]\leq \frac{L\beta^2NK}{2}{\sum}_n{\sum}_{k=0}^{K-1}\xi_n^2\mathbb{E}\bigg[\Big\Vert\nabla F_n\nonumber\\&\left(\tilde{\boldsymbol{w}}_n^{k,t}\right)\Big\Vert^2\bigg]-\frac{K\beta}{2}\mathbb{E}\left[\left\Vert \nabla F\left(\tilde{\boldsymbol{w}}^t\right)\right\Vert^2\right]+\frac{N\beta^3}{2}{\sum}_n{\sum}_{k=0}^{K-1}\xi_n^2L_n^2\beta^2\nonumber\\&k{\sum}_{j=0}^{k-1}\mathbb{E}\left[\left\Vert\nabla F_n\left(\tilde{\boldsymbol{w}}_n^{j,t}\right)\right\Vert^2\right].
\end{alignat}\end{small}Finally, by summing up (\ref{theorem3_6}) form $t=0$ to $T-1$, we have\begin{small}\begin{alignat}{1}
	&\frac{1}{T}{\sum}_{t=0}^{T-1}\mathbb{E}\left[\left\Vert\nabla F\left(\tilde{\boldsymbol{w}}^{t}\right)\right\Vert^2\right]\!\leq\! \frac{2}{K\beta T}\left(\mathbb{E}\left[F\!\left(\tilde{\boldsymbol{w}}^0\right)\right]\!-\!\mathbb{E}\left[F\!\left(\tilde{\boldsymbol{w}}^T\right)\right]\right)\nonumber\\&+\!\frac{L\beta N}{T}{\sum}_{t=0}^{T-1}{\sum}_n{\sum}_{k=0}^{K-1}\!\xi_n^2\mathbb{E}\left[\left\Vert\nabla F_n\!\left(\!\tilde{\boldsymbol{w}}_n^{k,t}\!\right)\right\Vert^2\right]\!+\!\frac{N\beta^2}{KT}\!{\sum}_{t=0}^{T-1}\nonumber\\&{\sum}_n{\sum}_{k=0}^{K-1}\xi_n^2L_n^2\beta^2k{\sum}_{j=0}^{k-1}\mathbb{E}\left[\left\Vert\nabla F_n\left(\tilde{\boldsymbol{w}}_n^{j,t}\right)\right\Vert^2\right].
\end{alignat}\end{small}This completes the proof of \textbf{Theorem} \ref{theorem3}.

\ifCLASSOPTIONcaptionsoff
\newpage
\fi
\bibliographystyle{IEEEtran}
\bibliography{references}

\end{document}